\documentclass[final,3p,times,twocolumn,sort&compress]{elsarticle}

\usepackage{lineno,hyperref}
\modulolinenumbers[5]

\usepackage{amsmath}
\usepackage{amsfonts,color}
\usepackage{mathtools}
\usepackage[utf8]{inputenc}
\usepackage{graphicx}
\usepackage{wasysym}
\usepackage{tabularx}
\usepackage{accents}
\usepackage{graphicx}
\usepackage{caption}
\usepackage{booktabs}
\usepackage{color, colortbl}
\usepackage[dvipsnames]{xcolor}
\definecolor{Gray}{gray}{0.8}
\definecolor{LightCyan}{rgb}{0.88,1,1}
\definecolor{LightPink}{rgb}{1,0.85,0.85}
\usepackage{hyperref}
\usepackage{ulem}
\usepackage{tikz}
\usetikzlibrary{positioning}
\usetikzlibrary{shapes.geometric, arrows}

\hypersetup{
    colorlinks=true,
    linkcolor=blue,
    filecolor=magenta,      
    citecolor=red
}

\usepackage{subfigure} 
\usepackage{hyperref} 

\usepackage{subfigure} 
\usepackage{multicol}
\usepackage{multirow}
\usepackage{algorithm,algorithmic}
\usepackage{tikz}

\newcommand{\be}{\begin{equation}}
\newcommand{\ee}{\end{equation}}
\newcommand{\bea}{\begin{eqnarray}}
\newcommand{\eea}{\end{eqnarray}}

\tikzstyle{startstop} = [rectangle, rounded corners, 
    text width=7em, minimum height=1cm,text centered, draw=black, fill=red!30]
\tikzstyle{io} = [trapezium, trapezium left angle=70, trapezium right angle=110, 
    text width=7em, minimum height=1cm, text centered, draw=black, fill=blue!30]
\tikzstyle{process} = [rectangle, 
    text width=7em, minimum height=1cm, text centered, draw=black, fill=orange!30]
\tikzstyle{decision} = [diamond, minimum height=1cm, text centered, text width=5.5em, node distance=3cm, draw=black, fill=green!30]
\tikzstyle{arrow} = [thick,->,>=stealth]

\journal{}

\begin{document}

\begin{frontmatter}

\title{Nature-inspired optimization, the Philippine Eagle, \\ and cosmological parameter estimation}


\author[a]{Reginald Christian Bernardo\corref{mycorrespondingauthor}}
\cortext[mycorrespondingauthor]{Corresponding author}
\ead{reginald.bernardo@apctp.org}

\author[b]{Erika Antonette Enriquez}
\ead{eaenriquez@math.upd.edu.ph}

\author[b,c]{Renier Mendoza}
\ead{rmendoza@math.upd.edu.ph}

\author[d,e,f]{Reinabelle Reyes\corref{mycurrentaffil}}
\cortext[mycurrentaffil]{Current Affiliations: Philippine Space Agency, Bagumbayan, Quezon City 1800, Philippines and Research Center for Theoretical Physics, Central Visayan Institute Foundation, Jagna, Bohol 6308, Philippines}
\ead{reinareyes.rctp@gmail.com}

\author[b,c]{Arrianne Crystal Velasco}
\ead{acvelasco@math.upd.edu.ph}

\address[a]{Asia Pacific Center for Theoretical Physics, Pohang 37673, Korea}
\address[b]{Institute of Mathematics, University of the Philippines Diliman, Quezon City 1101, Philippines}
\address[c]{Computational Science Research Center, University of the Philippines Diliman, Quezon City 1101, Philippines}
\address[d]{National Institute of Physics, University of the Philippines Diliman, Quezon City 1101, Philippines}
\address[e]{Philippine Space Agency, Bagumbayan, Quezon City 1800, Philippines}
\address[f]{Research Center for Theoretical Physics, Central Visayan Institute Foundation, Jagna, Bohol 6308, Philippines}

\begin{abstract}
Precise and accurate estimation of cosmological parameters is crucial for understanding the Universe's dynamics and addressing cosmological tensions. In this methods paper, we explore bio-inspired metaheuristic algorithms, including the Improved Multi-Operator Differential Evolution scheme and the Philippine Eagle Optimization Algorithm (PEOA), alongside the relatively known genetic algorithm, for cosmological parameter estimation. Using mock data that underlay a true fiducial cosmology, we test the viability of each optimization method to recover the input cosmological parameters with confidence regions generated by bootstrapping on top of optimization. We compare the results with Markov chain Monte Carlo (MCMC) in terms of accuracy and precision, and show that PEOA performs comparably well under the specific circumstances provided. Understandably, Bayesian inference and optimization serve distinct purposes, but comparing them highlights the potential of nature-inspired algorithms in cosmological analysis, offering alternative pathways to explore parameter spaces and validate standard results.
\end{abstract}


\end{frontmatter}



\section{Introduction}
\label{sec:intro}

Methods for accurate and robust cosmological parameter estimation are essential for interpreting increasingly precise observational data \cite{Trotta:2008qt, Trotta:2017wnx, Liddle:2007fy, Ntampaka:2019udw, Medel-Esquivel:2023nov}. The standard $\Lambda$CDM model, while successful in describing the large-scale dynamics of the Universe \cite{Peebles:2002gy,Copeland:2006wr, Riess:1998cb,Perlmutter:1998np, Guth:1980zm,Linde:1981mu,Bertone:2004pz,Baudis:2016qwx}, relies on statistical inference techniques to extract meaningful constraints from the data. As measurements continue to improve, so too must the methods used to analyze them.

An excellent example of the importance of robust parameter estimation is the Hubble constant tension \cite{DiValentino:2020vhf,DiValentino:2020zio,DiValentino:2020vvd, Schoneberg:2021qvd, Riess:2021jrx, Aghanim:2018eyx}; clearly, one of contemporary cosmology's most pressing challenges. Local measurements and early Universe estimates have reached precision, sufficient to reveal a statistically significant disagreement between inferred cosmological parameters and draw scrutiny over both observational systematics and theoretical assumptions \cite{DiValentino:2025sru}. While addressing present and future cosmological tensions may require new physics \cite{Cardenas:2014jya, Vagnozzi:2019ezj, Grandon:2021nls, Bernardo:2021cxi, Capozziello:2011et, Peirone:2019aua, Frusciante:2018jzw, Bahamonde:2019shr, Bernardo:2021qhu, Odintsov:2022eqm}, it also hinges on refining statistical techniques to ensure that parameter constraints are accurate and unbiased. This has led to a growing interest in noncanonical statistical methods, see e.g., \cite{Escamilla-Rivera:2020fxq, Briffa:2020qli, LeviSaid:2021yat, Bernardo:2022ggl, Mukherjee:2022yyq, Benisty:2022psx, Grandon:2022gdr, Kim:2023unc, Gomez-Vargas:2024izm}.

{In this work, we primarily focus on the applicability of an alternative class of numerical techniques---nature-inspired, metaheuristic or evolutionary algorithms (EAs)---which are commonly used in optimization problems, for cosmological parameter estimation.} While Markov chain Monte Carlo (MCMC) methods remain the standard in cosmology and astrophysics \cite{Lewis:2019xzd, 2020arXiv200505290T}, EAs offer a complementary approach, capable of efficiently exploring high-dimensional, complex, and nonlinear parameter spaces. One such EA is genetic algorithm (GA), endowed with natural evolution operations, selection, crossover, and mutation \cite{Medel-Esquivel:2023nov, Bernardo:2025zbv}. GA is also widely used in cosmology in the context of model-independent cosmological reconstruction through grammatical evolution \cite{Bogdanos:2009ib, Nesseris:2010ep, Nesseris:2012tt, Arjona:2021mzf}. We also introduce two EAs, the Improved Multi-Operator Differential Evolution (IMODE) and the Philippine Eagle Optimization Algorithm (PEOA), and test their ability to recover cosmological parameters, with confidence regions provided by bootstrapping on top of optimization. It is worth emphasizing that both methods have been well established in mathematical optimization and have been used in a variety of applications, in some, outperforming GA \cite{imode, Enriquez_Mendoza_Velasco_2022} \footnote{We refer the interested readers to \cite{Enriquez_Mendoza_Velasco_2022} for a rigorous examination of GA, IMODE and PEOA, and other methods, to a variety of problems. We also recommend \cite{DiValentino:2025sru} for a short review of the applications of EAs in cosmology and their potential role in the future of the field.}. Through this work, we introduce them to cosmology.

IMODE is an enhanced version of Differential Evolution, which introduces multiple adaptive mutation and crossover operators to improve exploration and exploitation balance. It has shown success in complex, multimodal optimization problems, making it suitable for navigating cosmological data sets with hitherto unknown systematics. On the other hand, PEOA is inspired by the hunting strategies of the Philippine Eagle (Figure \ref{fig:PhEagleHunting}), with mechanisms for intelligent exploration of parameter spaces. Its adaptive nature allows for effective navigation of nonlinear and high-dimensional landscapes, and highlights distinctive characteristics of the critically endangered, national bird of the Philippines.

Adding a bootstrapping mechanism to estimate confidence intervals, we investigate the performance of IMODE and PEOA in cosmological parameter estimation with a oversimplified mock data (diagonal covariance), together with GA and MCMC as references. {We emphasize that the uncertainties using this approach approximate frequentist intervals obtained by resampling the data, and are not Bayesian. Most importantly, the bootstrap method is a natural path to estimate errors in optimization strategies such as EAs that do not come with their own uncertainty estimates.} This serves to draw a baseline for future more complicated, realistic analyses.

It is worth emphasizing that our present work focuses on the methods rather than addressing observational tensions or drawing cosmological consequences. We reserve the latter for a different paper. Instead, our view is shared with the community that the increasing complexity of cosmological data must be faced with a rich variety of statistical tools \cite{DiValentino:2025sru}; each with unique strengths in speed, accuracy, and sensitivity to systematic errors that can be utilized for robustness and cross-validation. By complementing traditional approaches with alternative methods, the limitations of any single method can be mitigated, enhancing the robustness of cosmological analyses. This reflects the Filipino concept of `Bayanihan' (cooperative effort for collective benefit).
Our results support this perspective, laying the groundwork for potentially incorporating EAs beyond GA into standard cosmological analysis.

We have to stress out that our comparison among methods should be taken figuratively, since each method has been tailored to address different mathematical situations. In addition, EAs are optimization methods and we are only able to provide error estimates by laying a bootstrapping mechanism besides optimization. On the other hand, MCMC directly approximates the posterior. In view of optimization, comparison between methods should be carefully framed within the context of specific problems or the data used, since generally no one method can be deemed better than others.

The remainder of this work is organized as follows. In Section \ref{sec:lcdm_model}, we provide a brief review of the standard $\Lambda$CDM model and the key observables relevant to our analysis. Section \ref{sec:mock_data} discusses the generation of our mock data sets. In Section \ref{sec:evolutionary_algorithm}, we give an overview of MCMC before detailing the EAs central to this study: GA, IMODE, and PEOA. Our results are presented in Section \ref{sec:results}, followed by a discussion in Section \ref{sec:discussion} that addresses the strengths and limitations of each method in the context of cosmological parameter estimation. Appendix \ref{sec:pseudo-algorithms} gives pseudo-algorithms for GA, IMODE, and PEOA. Our codes implementing IMODE and PEOA can be found in \href{https://github.com/ErikaAntonette/Philippine-Eagle-Optimization-Algorithm}{GitHub}. {Our analysis codes in MATLAB (Global Optimization Toolbox, AstroPack, IMODE, PhEagleOA) and python (emcee) are shared in a dedicated GitHub repository.\footnote{
\href{https://github.com/reggiebernardo/pheagle_cosmo}{https://github.com/reggiebernardo/pheagle\_cosmo}
}}

\section{The $\Lambda$CDM model}
\label{sec:lcdm_model}

The standard cosmological model, often referred to as $\Lambda$CDM, provides a physical framework for understanding the large scale structure and dynamics of the Universe. It stitches together a cosmological constant, denoted by $\Lambda$, a Cold Dark Matter (CDM) component, and the cosmological principle to explain the observed cosmic expansion and the formation of structures such as galaxies and clusters. Additionally, it integrates with general relativity to describe the behavior of the Universe at smaller scales where nonlinearities become relevant due to gravitational collapse.

Through $\Lambda$CDM, we understand that the cosmic expansion and acceleration at sufficiently large scales are governed approximately by a simple set of equations; that admit the following analytical expression for the Hubble function,
\begin{equation}
\label{eq:hubble_function}
    H\left(z \right) = H_0 \sqrt{ \Omega_{m0} \left(1 + z \right)^3 + \left(1 - \Omega_{m0}\right) } \,.
\end{equation}
Above, $z$ is the cosmological redshift, $H_0$ is the Hubble constant, and $\Omega_{m0}$ is the fraction of nonrelativistic matter (baryons and CDM) at redshift $z = 0$, or the present cosmic time. Provided spatial flatness, the luminosity distance can be written out as
\begin{equation}
\label{eq:luminosity_distance}
    d_L(z) = \dfrac{c}{H_0} \left(1 + z\right) \int_0^z \dfrac{dz'}{E(z')} \,,
\end{equation}
where $E(z) = H(z)/H_0$ is the normalized expansion rate. At very low redshifts, $z \ll 1$, this reduces to the `Hubble law', $d_L \sim cz / H_0$; indicating that nearby galaxies are recessing with a speed, $v_r/c \sim z$, proportional to their distance $d_L$. Likewise, the matter density perturbation can be teased out  analytically as \cite{Nesseris:2015fqa}
\begin{equation}
\label{eq:matterpert_lcdm}
    \delta(a) = a \,_2F_1\left( \dfrac{1}{3},1, \dfrac{11}{6}; a^3 \left(1 - \Omega_{m0}^{-1}\right)  \right) \,,
\end{equation}
where $\,_2F_1\left(a,b,c;x\right)$ is the hypergeometric function, and $a$ is the scale factor, related to the redshift via $a = 1/(1 + z)$. A general analytical expression exists for an arbitrary constant dark energy equation of state \cite{Nesseris:2015fqa}. These are the linearized solutions to the Einstein equation in a flat {{Friedmann-Lema\'itre-Robertson-Walker}} background plus ad hoc elements of a constant dark energy and CDM. Remarkably, the Universe appears to be fairly consistent with the standard cosmological model.

The speed of light in vacuum is $c \simeq 299792.458$ km/s and the Hubble constant can be expressed as $H_0 = h \ \times$ (100 km s$^{-1}$ Mpc$^{-1}$) where $h$ is dimensionless. In practice, it is useful to write down $c/H_0 \simeq 2997.92458 \ {\rm Mpc}/h$ to express cosmological distances such as $d_L(z)$ in megaparsecs.

\section{Mock cosmological data}
\label{sec:mock_data}

In this section, we consider a $\Lambda$CDM model with $H_0 = 70$ km s$^{-1}$ Mpc$^{-1}$, $\Omega_{m0} = 0.3$, and $\sigma_{8} = 0.8$ (amplitude of matter power spectrum, to be introduced shortly) to generate mock data sets for testing IMODE and PEOA as well as MCMC and GA.

We generate a realization of mock data $\left(z_{\rm mock}, y_{\rm mock} \pm \Delta y_{\rm mock} \right)$ based on actual ones $\left(z_{\rm act}, y_{\rm act} \pm \Delta y_{\rm act}\right)$ and a `true' cosmology $(z, y_{\rm true})$; using the following series of steps:
\begin{enumerate}
    \item Fix the redshifts to match observed ones, $z_{\rm mock} = z_{\rm actual}$;
    \item Draw the mean from normal distribution around the true cosmology with a variance fixed by the observations, $y_1 = {\cal N} \left( y_{\rm true}, \left( \Delta y_{\rm act} \right)^2 \right)$;
    \item Final mock data $\left(z_{\rm mock}, y_{\rm mock} \pm \Delta y_{\rm mock} \right)$ is $\left(z_{\rm act}, y_1 \pm \Delta y_{\rm act} \right)$.
\end{enumerate}
This leads to a very conservative mock data, with redshifts and measurement error bars fixed by real observations, to avoid possible spurious forecasts. We also emphasize that for assessing the performance of EAs in the background of MCMC, we found that this methodology is sufficient for our purposes, and supported by the results. But note that this does not use a full covariance; which we shall return to in future analysis.

For the `true' cosmology, we consider a $\Lambda$CDM model with the parameters $(h, \Omega_{m0}, \sigma_{8}) = (0.7, 0.3, 0.8)$ where $h = H_0 / 100 \ {\rm km \ s}^{-1}{\rm Mpc}^{-1}$ and $\sigma_8$ is the average matter power spectrum on scales of $8 h^{-1} {\rm Mpc}$. The statistical methods are thus tasked to infer the true cosmology with simulated data.

\begin{figure}[h!]
    \centering
    \includegraphics[width=0.485\textwidth]{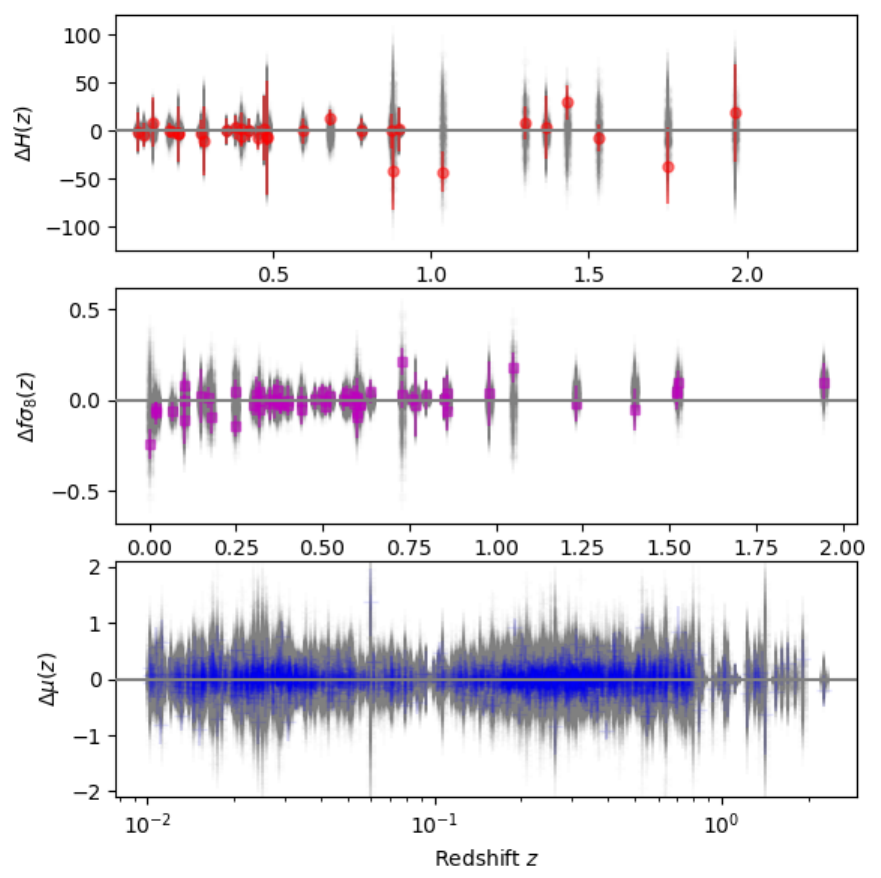}
    \caption{Residuals, $\Delta y = y_{\rm mock}-y_{\rm true}$, in one (colored) and a hundred (gray) realisations of mock data of expansion rate, $H(Z)$, growth rate, $f \sigma_8(Z)$, and supernovae distance-modulus, $\mu(Z)$; generation process described in Section \ref{sec:mock_data}; horizontal line at $y=0$ is obtained from fiducial $\Lambda$CDM model ($\Omega_{m0}=0.3$, $H_0=70$ km s$^{-1}$Mpc$^{-1}$, $\sigma_{8}=0.8$).}
    \label{fig:mock_data_i}
\end{figure}

For the basis of our mock data, we consider background cosmological data from cosmic chronometers (CC) \cite{2010JCAP...02..008S, 2012JCAP...08..006M, 2014RAA....14.1221Z, Moresco:2015cya, Moresco:2016mzx, Ratsimbazafy:2017vga} and supernovae (SNe) \cite{Brout:2021mpj, Brout:2022vxf, Scolnic:2021amr} through the Pantheon+ compilation. The expansion rate data considered in this work can be found in the Appendix of \cite{Bernardo:2021cxi}. In addition, we consider the growth rate data from redshift space distortion (RSD) measurements compiled in \cite{Kazantzidis:2018rnb} to add a layer of linear perturbations to our analysis. To see that the mocks are aligned with the input cosmology, we calculate the residuals ($\Delta y = y_{\rm mock}-y_{\rm true}$) in one and multiple realisations of the mock data, as shown in Figure \ref{fig:mock_data_i}. This shows that the mock data can deviate from the true cosmology in one simulation, but this is consistent with the expectation when multiple realisations are taken into account. Note the residuals must show up as white noise with zero mean, if the mock data in multiple realisations are stacked together. Figure \ref{fig:mock_data_i} supports this. We give a brief description of the observations considered in the following paragraphs.

The CC are standardized clocks, providing an arguably cosmological model-independent probe to the cosmic expansion. This comes from observations of age and metal{l}icity of temporally adjacent, passively evolving galaxies, such that the expansion rate at a redshift $z_{\rm CC}$ can be approximated by $H_{\rm CC}(z) \approx -\left(\Delta z_{\rm CC}/\Delta t\right)/\left(1 + z_{\rm CC}\right)$. Thus, CC at different redshifts can be regarded as a direct accounting of late-time expansion history. Our studied CC consists of redshifts $0.07 \lesssim z \lesssim 1.97$ \cite{2010JCAP...02..008S, 2012JCAP...08..006M, 2014RAA....14.1221Z, Moresco:2015cya, Moresco:2016mzx, Ratsimbazafy:2017vga}.

The SNe Type Ia are widely considered standard candles, largely because of their high intrinsic brightness. This works on the wit that, at night, one can determine the distance and speed of a standard car by measuring the brightness of its head lights, and understanding that the intensity of radiation drops according to the inverse-square-law. In the same vein, SNe Type Ia serve as cosmic lampposts that give away their cosmological distances after standardization. Flux observations of SNe together with redshift measurements enable a construction of the distance--ladder, in symbols,
\begin{equation}
\label{eq:distmod_def}
    \mu(z) = 5 \log_{10}\left( \dfrac{ d_L\left(z\right) }{ 1 \ {\rm Mpc} } \right) +  25 \,,
\end{equation}
where $d_L(z)$ is the luminosity distance given by Eq. \eqref{eq:luminosity_distance} in $\Lambda$CDM\footnote{
Using standard cosmological units, we can explicitly write this down as $\mu(z) = 5 \log_{10} \left( \left(2997.92458/h\right) \times (1+z) \int_0^z dz' / E\left(z'\right) \right) + 25$ where $h = H_0/ \left( 100 \ {\rm km} \ {\rm s}^{-1} \ {\rm Mpc}^{-1} \right)$ is a dimensionless surrogate to the Hubble constant. A typical $h \sim 0.7$, corresponding to an expansion rate $H_0 \sim 70 \ {\rm km} \ {\rm s}^{-1} \ {\rm Mpc}^{-1}$.}. For the basis of our mock data sets, we use the Pantheon$+$ compilation \cite{Brout:2021mpj, Brout:2022vxf, Scolnic:2021amr} of distance--ladder measurements from SNe Ia; we only take conservative redshift range $0.01 \lesssim z \lesssim 2.3$, dropping ones for $z \ll 0.01$, to avoid issues brought up in \cite{Pasten:2023rpc, Perivolaropoulos:2023iqj}.

{We must emphasize that SNe data constrains only relative distances and requires an additional calibration of their absolute magnitude or an external anchor to estimate luminosity distances. On the other hand, our simplified mock data will be based on theoretically calibrated SNe since the work mainly concerns setting a baseline comparison of EAs and MCMC. However, this gives unrealistic (sub-percent) precisions for cosmological parameters, such as the Hubble constant, which we warn should not be overthought.}

RSD measurements utilize anisotropies in a galaxy distribution due to the differences between galaxy peculiar velocities and the Hubble flow. This keys in crucial cosmological information on the growth of structure, understood through the perturbation, $\delta(z)$, and its subsequent evolution. RSD growth data is given as $f \sigma_8(z)$, a product of the growth rate $f(z) = d \ln \delta (z)/d \ln a(z)=-(1+z)\delta'(z)/\delta(z)$ and the 
root-mean-square matter density fluctuation in a sphere of radius$\sim$8 Mpc, $\sigma_8(z) = \sigma_8 \delta (z)/ \delta(z=0)$. In the $\Lambda$CDM model, the matter density perturbation is given by Eq. \eqref{eq:matterpert_lcdm}; therefore, the growth rate can be expressed as \cite{LeviSaid:2021yat}
\begin{equation}
\label{eq:growth_lcdm}
    f \sigma_8(z) = - \sigma_{8} (1 + z) \dfrac{\delta'(z)}{\delta(z = 0)} \,,
\end{equation}
where $\sigma_8$ is the amplitude of the matter power spectrum, setting the scale of perturbations. Note that the density fluctuation/constrast can be explicitly written in terms of the redshift, $z = a^{-1} - 1$, in $\Lambda$CDM as \begin{equation}
\delta(z) = \dfrac{1}{1+z} \,_2F_1\left( \frac{1}{3},1,\frac{11}{6}; \dfrac{\left(1 - \Omega_{m0}^{-1}\right)}{(1+z)^3} \right) \,.    
\end{equation}
This can be differentiated analytically to substitute into Eq. \eqref{eq:growth_lcdm} to efficiently numerically evaluate $f\sigma_8(z)$ for comparison with simulated data.

\section{MCMC, GA, and evolutionary algorithms}
\label{sec:evolutionary_algorithm}

We briefly review MCMC (Section \ref{subsec:mcmc}) and describe in detail GA (Section \ref{subsec:ga}), IMODE (Section \ref{subsec:imode}), and PEOA (Section \ref{subsec:pea}).

\subsection{MCMC}
\label{subsec:mcmc}

Given a data D and parameters $p$ of a model M, Bayes' theorem states that
\begin{equation}
\label{eq:bayes_theorem}
    P(p | {\rm D, M}) = \dfrac{P({\rm D} | p, {\rm M}) P(p | {\rm M})}{P({\rm D|M} )},
\end{equation}
where $ P(p | {\rm D, M}) $ (the posterior) is the probability distribution of the parameters $p$ given D and M, $ P({\rm D} | p, {\rm M}) $ (the likelihood) is the probability of observing the data D given $p$ and M, $P(p | {\rm M})$ (the prior) represents the prior probability of the parameters \( p \), and $P({\rm D | M})$ (the evidence or marginal likelihood of M) is a normalizing constant. In this work, we are focused on estimating the posterior distribution $P(p | {\rm D, M})$, and will not be concerned with model comparison. This means that the evidence $ P({\rm D | M})$ can be effectively treated as a constant.

For our purposes, we utilize the python package \texttt{emcee} \cite{emcee} to perform MCMC, and obtain posterior estimates of the cosmological parameters given mock data sets. The likelihoods are given as follows, based on the standard $\Lambda$CDM model (Section \ref{sec:lcdm_model}). For the expansion rate, we compare the model to the data using a likelihood ${\cal L}$ given by
\begin{equation}
\label{eq:loglike_cc}
\log {\cal L} \propto - \dfrac{1}{2} \sum_{z_{\rm D}} \left( \dfrac{H_{\rm M}\left(z_{\rm D}\right) - H_{\rm D}(z_{\rm D})}{ \Delta H_{\rm D}(z_{\rm D})} \right)^2 \,,
\end{equation}
where the quantities with the subscript D refer to the data, and ones with M refer to the model, i.e., $H_{\rm M}\left(z_{\rm D}\right)$ are the expansion rate prediction of a model M at the corresponding observed redshifts $z_{\rm D}$. Similarly, for SNe observations, we consider the likelihood function
\begin{equation}
\label{eq:loglike_sne}
    \log {\cal L} \propto -\dfrac{1}{2} \sum_{z_{\rm D}} \left( \dfrac{ \mu_{\rm M}(z_{\rm D}) - \mu_{\rm D}(z_{\rm D}) }{ \Delta \mu_{\rm D}(z_{\rm D}) } \right)^2 \,,
\end{equation}
and for growth rate data, we take the likelihood
\begin{equation}
\label{eq:loglike_rsd}
\begin{split}
    \log {\cal L} \propto - \dfrac{1}{2} \sum_{z_{\rm D}}
    & \left( \dfrac{ f \sigma_{8, {\rm M}}(z_{\rm D}) - f \sigma_{8, {\rm D}}(z_{\rm D}) }{ \Delta f \sigma_{8, {\rm D}}(z_{\rm D}) } \right)^2 \,.
\end{split}
\end{equation}
{In addition to the above likelihoods (\ref{eq:loglike_cc}-\ref{eq:loglike_rsd}), we consider the flat/uniform unit cube priors on $(h, \Omega_{m0}, \sigma_8)$, as shown in Table \ref{tab:mcmc_priors}, to estimate the posterior using MCMC through Bayes' theorem \eqref{eq:bayes_theorem}. Furthermore, we implement the sampling routine starting at a maximum of the likelihood (obtained using Nelder-Mead local search) with a total of 60,000 steps and 100 burn-in steps. Readers interested can view our implementation in GitHub.}
\begin{table}[ht]
\captionof{table}{{MCMC priors used in this work, where $h=H_0/\left({100 \,{\rm km}\,{\rm s}^{-1}{\rm Mpc}^{-1}}\right)$ where $H_0$ is the Hubble constant.}}
    \centering
    \begin{tabular}{@{}cc@{}}
        \toprule
        Parameter & Prior Values \\ \midrule
        $h$ & $[0, 1]$ \\ 
        $\Omega_{m0}$ & $[0, 1]$ \\
        $\sigma_{8}$ & $[0, 1]$ \\ \bottomrule
    \end{tabular}
    \label{tab:mcmc_priors}
\end{table}

As explained previously, for the purposes of this work, our mock data sets do not take into account the full covariance of the observed data\footnote{The covariance of real data can be accounted for by replacing the weights in (\ref{eq:loglike_cc}-\ref{eq:loglike_rsd}) by the inverse of the covariance matrix.}.

\subsection{Genetic Algorithm}
\label{subsec:ga}

\begin{figure*}[h!]
	
	\tikzstyle{decision} = [diamond, draw, fill=blue!20, text width=10em, text centered, inner sep=4pt, fill=green!30, aspect=2]
	\tikzstyle{cloud} = [rectangle, draw, fill=blue!20, text width=5em, text centered, rounded corners=12pt, minimum height=2em]
	\tikzstyle{line} = [draw, -latex']
	\tikzstyle{block} = [rectangle, draw, fill=blue!20, text width=18em, text centered, minimum height=3.5em]
	\tikzstyle{put} = [trapezium, draw, text width=10.5em, fill=blue!20, minimum height=2em, trapezium left angle=50, trapezium right angle=130, text centered]
	\tikzstyle{block2} = [rectangle, draw, fill=blue!20, text width=15em, text centered, minimum height=3.5em]
	\tikzstyle{startstop} = [rectangle, rounded corners, minimum width=3cm, minimum height=1cm,text centered, draw=black, fill=red!30]
	\tikzstyle{io} = [trapezium, trapezium left angle=70, trapezium right angle=110, minimum width=0cm, minimum height=0cm, text centered, draw=black, fill=blue!30,  text width=12em]
	\tikzstyle{process} = [rectangle, minimum width=2cm, minimum height=1cm, text centered, text width=10.0em, draw=black, fill=orange!30]
	\tikzstyle{process2} = [rectangle, minimum width=3cm, minimum height=1cm, text centered, text width=15.8em, draw=black, fill=white!30]
	\tikzstyle{decision2} = [diamond, minimum width=3cm, minimum height=1cm, text centered, draw=black, fill=green!30, aspect=2]
	\tikzstyle{arrow} = [thick,->,>=stealth]
	\tikzstyle{dummy} = [rectangle, draw, minimum width=0cm, minimum height=0cm,text centered, draw=white, fill=white]
	
	\begin{center}
		\resizebox{0.65\textwidth}{!}{
			\begin{tikzpicture}[thick, every node/.style={scale=0.75},node distance = 2.5cm, auto]
				\node [startstop] (begin) {\large Begin};
				\node [io, right of=begin,node distance=5cm] (in1) {\large Input:\\objective function $f$, dimension $D$, search space bounds $X_{\textrm{min}}$, $X_{\textrm{max}}$, crossover rate $cr$, archive size $m$};
				\node [process, right of=in1,node distance=6cm] (init) {\large Set current population $k=1$, maximum function evaluations $Max\text{-}FES$, initial population size $PS$};
				\node [process, right of=init,node distance=5.5cm] (eval) {\large Evaluate $f(X)$ and sort};
				\node [decision, below of=eval, node distance=5cm] (decidewhile) {\large $k\leftarrow k+1$ $FES \leq Max\text{-}FES$?};
				\node [process, left of=decidewhile,node distance=7cm] (popreduction) {\large Choose the best $m$ individuals to form the elite pool.};
				\node [process, left of=popreduction, node distance=7cm] (crossover) {\large Apply a tournament selection with size $TC$ and fill the selection pool. Select any two individuals (parents) from the pool and generate an offspring with crossover rate $cr$};
				\node [process, below of=crossover,node distance=5cm] (mutate) {\large Mutate the offspring.};
				\node [process, right of=mutate,node distance=7cm] (updatepop) {\large Generate the new population by combining the elite and the pool of offsprings, maintaining the population size};
				\node [process, below of=updatepop,node distance=3cm] (sort) {\large Sort new population};
				\node [io, below of=sort, node distance=3cm] (out) {\large Output:\\best solution};
				\node [startstop, left of=out, node distance=6cm] (end) {\large End};
				
				\coordinate (point1) at (-2.25cm, -3.75cm);
				\coordinate (point2) at (-2.25cm, -9cm);
				\coordinate (point3) at (4.875cm, -6.25cm);
				\coordinate (point4) at (12.375cm, -6.25cm);
				\coordinate (point5) at (15.25cm, -3.75cm);
				\coordinate (point6) at (15.25cm, -12.05cm);
				\draw [arrow] (begin) -- (in1);
				\draw [arrow] (in1) -- (init);
				\draw [arrow] (init) -- (eval);
				\draw [arrow] (eval) -- (decidewhile);
				\draw [arrow] (decidewhile) -- node [left,xshift=1.5em,yshift=1em] {Yes} (popreduction);
				\draw [arrow] (popreduction) -- (crossover);
				\draw [arrow] (crossover) -- (mutate);
                \draw [arrow] (mutate) -- (updatepop);
				\draw [arrow] (updatepop) -- (sort);
				\draw [arrow] (sort) -| (decidewhile);
				\draw (decidewhile) -- node [above,xshift=0em,yshift=0.125em] {No}(point5);
				\draw (point5) -- (point6);
				\draw [arrow] (point6) -- (out);
				\draw [arrow] (out) -- (end);
				
		\end{tikzpicture}}
		\caption{Flowchart illustrating the key steps in GA.}
		\label{fig:GA_flowchart}
	\end{center}
\end{figure*}
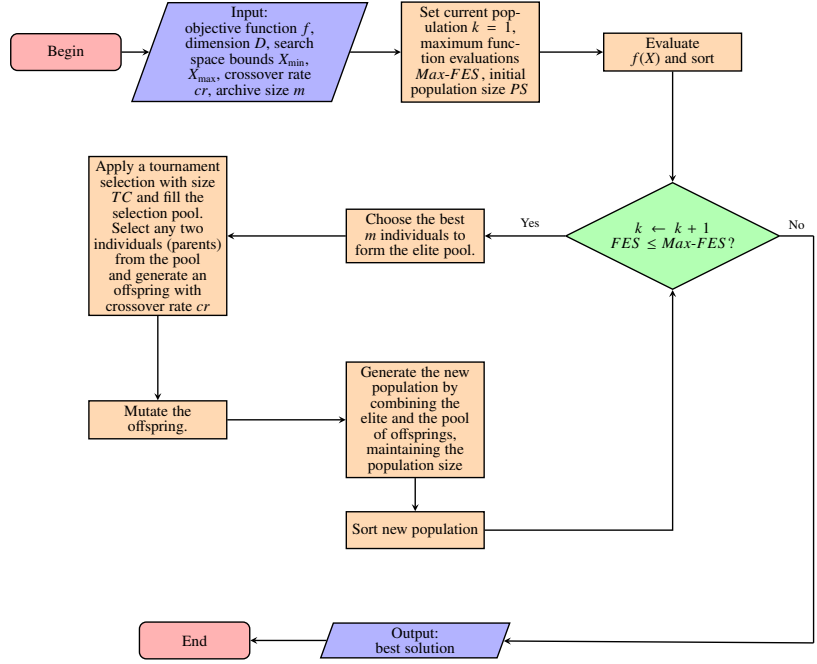

The GA is part of a broader EAs, which aim to optimize the quality of a population of solutions rather than focusing solely on individual solutions. GA has found diverse applications, including in high-energy physics and gravitational wave astronomy \cite{Akrami:2009hp, Crowder:2006wh}, and was initially introduced in cosmology as a tool to mitigate biases in model selection when analyzing dark energy \cite{Bogdanos:2009ib}. Subsequent studies have demonstrated GA's effectiveness in estimating cosmological parameter uncertainties \cite{Nesseris:2010ep, Nesseris:2012tt}, selecting kernel functions in Gaussian process regression \cite{Bernardo:2021mfs}, optimizing neural network architectures \cite{Gomez-Vargas:2022bsm}, enhancing spectroscopic modeling \cite{Bainbridge2017a, Bainbridge2017b, Lee2020AI-VPFIT}, and aiding model selection with information criteria \cite{Webb2021}. For a recent overview of GA in cosmology, we refer our readers to \cite{Medel-Esquivel:2023nov, DiValentino:2025sru, Bernardo:2025zbv}, and for further applications, to \cite{GA_app1, GA_app2}.

GA works by iteratively evolving a population of candidate solutions through cycles of fitness evaluation, selection, and breeding---each cycle being known as a generation (Figure \ref{fig:GA_flowchart}). In each generation, GA assesses the `fitness' of each individual (its quality or suitability for solving the problem) and selects the fittest candidates to become parents. Breeding combines these selected individuals, introducing diversity through mutation (a slight alteration of an individual) and recombination or crossover (a process that creates offspring by combining traits from two parents). This interplay between selection and variation allows the population to explore the solution space, refining solutions with each generation.

The process begins with population initialization, where $n$ unique individuals are generated randomly across the solution space; although they can be concentrated in promising regions if prior knowledge exists. Selection then chooses the fittest individuals as parents, favoring those with higher fitness scores. These parents undergo recombination, involving a crossover operation that merges traits from each parent to form offspring. Crossover captures the idea that fitter individuals share advantageous traits, while mutation—often implemented as Gaussian noise (zero mean and variance $\sigma^2$) introduces small random changes, helping escape local optima. GA typically uses tournament selection: it picks the fittest from a subset of $t$ randomly chosen individuals, balancing competition and randomness.

A common enhancement to GA (`elitism') preserves the best solutions by automatically including the fittest individuals from the current population in the next. These elites ensure that solution quality only improves, allowing GA to focus on refinement without risking the loss of highly fit individuals. Figure \ref{fig:GA_flowchart} shows the flowchart of GA.

{GA, mathematically, is a global optimization method that is specifically designed to overcome shortcomings of classical search routines such as prior dependencies and tendency to become trapped in local minima \cite{10.7551/mitpress/1090.001.0001, Katoch2021, Mirjalili2019, Katsifarakis2020, Thompson2024}. Additionally, GA can be viewed as a population method; where a population ${\bf p}$ is acted upon by a set of probabilistic operators ${\bf O}$, until a termination criterion is met, or the maximum number of generations is reached. The population is ranked according to their fitness $f$ and the random operators are selection ${\bf O}_{\rm Sel}$, crossover ${\bf O}_{\rm Cr}$, and mutation ${\bf O}_{\rm Mut}$, each of which are a function of their hyperparameters such as the mutation type and probability \cite{SCHMITT20011, 2011arXiv1105.3538W}. GA mimics natural selection through the action of an operator that can be viewed macroscopically through the population or microscopically via the genetic level, \textit{e.g.}, in our context an individual would be a set of cosmological parameters $(H_0, \Omega_{m0}, \sigma_{8})$ and the action of an operator $(H_0, \Omega_{m0}, \sigma_{8}) \rightarrow (H_0', \Omega_{m0}', \sigma_{8}')$. Intrinsically, each operator can be thought of as applying a probability density function to an individual's genetic make-up to keep diversity at the same time looking for possible improvements in future generations. Then, the population after the $k$th generation is \cite{10.7551/mitpress/1090.001.0001}
\begin{equation}
{\mathbf p}_{k+1}(f_{k+1})={\mathbf O}_{\rm Mut} {\mathbf O}_{\rm Cr} {\mathbf O}_{\rm Sel} {\mathbf p}_n(f_n) \,,
\end{equation}
and the final evolved population can be written as a successive application of the operators: 
\begin{equation}
{\mathbf p}_K(f_K)=\left({\mathbf O}_{\rm Mut} {\mathbf O}_{\rm Cr} {\mathbf O}_{\rm Sel}\right)^N {\mathbf p}_1(f_1)
\end{equation}
where ${\mathbf p}_1$ is the first (prior) population distribution. The population can be expected to cluster toward the solution in the limit when $K$ becomes increasingly large.}

We employ the built-in \texttt{ga} command in \texttt{MATLAB}, using its default settings for unconstrained problems. The population size is initially set to 50, and 5\% of the population is preserved as an elite pool. For parent selection, the algorithm assigns each candidate a segment of a line proportional to its scaled fitness value and selects parents by stepping uniformly along this line. Crossover is performed by randomly combining genes from two parents based on a binary mask, with a crossover rate of 80\%. Mutation is handled by perturbing each gene by adding Gaussian noise with mean zero. The standard deviation of this noise is initialized based on the population range and gradually decreases over generations. {Table \ref{tab:GA}  presents the default parameter values of the \texttt{ga} solver in MATLAB.}
\begin{table}[h!]
\captionof{table}{{The default values of the hyperparameters of GA that are used in the simulations.}}
\centering\resizebox{.45\textwidth}{!}{%
\begin{tabular}{@{}cc@{}}
\toprule
Parameter                             & Type/Value              \\ \midrule
dimension                              &          3          \\
maximum number of function evaluations &            500        \\
maximum number of generations          &          10          \\
population size                        &          50          \\
selection                              & stochastic uniform \\
crossover                                   & scattered         \\
crossover rate                                   & 0.8         \\
mutation                               & Gaussian           \\
elitism percentage                                  & 0.05         \\ \bottomrule
\end{tabular}}
\label{tab:GA}
\end{table}

\subsection{Improved Multi-Operator Differential Evolution}
\label{subsec:imode}

\begin{figure*}[!ht]
	
	\tikzstyle{decision} = [diamond, draw, fill=blue!20, text width=10em, text centered, inner sep=4pt, fill=green!30, aspect=2]
	\tikzstyle{cloud} = [rectangle, draw, fill=blue!20, text width=5em, text centered, rounded corners=12pt, minimum height=2em]
	\tikzstyle{line} = [draw, -latex']
	\tikzstyle{block} = [rectangle, draw, fill=blue!20, text width=18em, text centered, minimum height=3.5em]
	\tikzstyle{put} = [trapezium, draw, text width=10.5em, fill=blue!20, minimum height=2em, trapezium left angle=50, trapezium right angle=130, text centered]
	\tikzstyle{block2} = [rectangle, draw, fill=blue!20, text width=15em, text centered, minimum height=3.5em]
	\tikzstyle{startstop} = [rectangle, rounded corners, minimum width=3cm, minimum height=1cm,text centered, draw=black, fill=red!30]
	\tikzstyle{io} = [trapezium, trapezium left angle=70, trapezium right angle=110, minimum width=0cm, minimum height=0cm, text centered, draw=black, fill=blue!30,  text width=12em]
	\tikzstyle{process} = [rectangle, minimum width=2cm, minimum height=1cm, text centered, text width=10.0em, draw=black, fill=orange!30]
	\tikzstyle{process2} = [rectangle, minimum width=3cm, minimum height=1cm, text centered, text width=15.8em, draw=black, fill=white!30]
	\tikzstyle{decision2} = [diamond, minimum width=3cm, minimum height=1cm, text centered, draw=black, fill=green!30, aspect=2]
	\tikzstyle{arrow} = [thick,->,>=stealth]
	\tikzstyle{dummy} = [rectangle, draw, minimum width=0cm, minimum height=0cm,text centered, draw=white, fill=white]
	
	\begin{center}
		\resizebox{0.65\textwidth}{!}{
			\begin{tikzpicture}[thick, every node/.style={scale=0.75},node distance = 2.5cm, auto]
				\node [startstop] (begin) {\large Begin};
				\node [io, right of=begin,node distance=5cm] (in1) {\large Input:\\objective function $f$, dimension $D$, search space bounds $X_{\textrm{min}}$, $X_{\textrm{max}}$};
				\node [process, right of=in1,node distance=6cm] (init) {\large Set $G=1$, $ls=0.1$, maximum function evaluations $Max\text{-}FES$, initial population size $NP_G$};
				\node [process, right of=init,node distance=5.5cm] (eval) {\large Evaluate $f(X)$ and sort};
				\node [decision, below of=eval, node distance=5cm] (decidewhile) {\large $G\leftarrow G+1$ $FES \leq Max\text{-}FES$?};
				\node [process, left of=decidewhile,node distance=6cm] (popreduction) {\large Update $NP_G$ and assign the number of solutions $NP_{op}$ to each operator};
				\node [process, left of=popreduction,node distance=5.5cm] (op2) {\large Operator 2: \emph{current-to-$\phi$best without archive}};
				\node [process, above of=op2,node distance=2cm] (op1) {\large Operator 1: \emph{current-to-$\phi$best with archive}};
				\node [process, below of=op2,node distance=2cm] (op3) {\large Operator 3: \emph{weighted-rand\\-to-$\phi$best}};
				\node [process, left of=op2,node distance=5.5cm] (updatepop) {\large Generate the new population based on the operators and update the archive};
				\node [process, below of=updatepop,node distance=7cm] (updatemember) {\large Calculate $NP_{op}$ based on solutions' quality and diversity};
				\node [process, below of=updatemember,node distance=7cm] (sort) {\large Sort new population};
				\node [decision, right of=sort, node distance=7cm] (decidels) {\large $FES>0.85\,Max\text{-}FES$ and $rand<ls$?};
				\node [process, above of=decidels,node distance=4cm] (localsearch) {\large Local search from the best point in current population};
				\node [decision, above of=localsearch, node distance=4cm] (comparels) {\large Did the local search improve the solution?};
				\node [process, right of=comparels,node distance=7cm] (updatebest) {\large Update the best solution\\$ls\leftarrow 0.1$};
				\node [process, above of=updatebest,node distance=2.675cm] (updatels) {\large $ls \leftarrow 0.01$};
				\node [io, below of=decidels, node distance=3cm] (out) {\large Output:\\best solution};
				\node [startstop, left of=out, node distance=6cm] (end) {\large End};
				
				\coordinate (point1) at (-2.25cm, -3.75cm);
				\coordinate (point2) at (-2.25cm, -9cm);
				\coordinate (point3) at (4.875cm, -6.25cm);
				\coordinate (point4) at (12.375cm, -6.25cm);
				\coordinate (point5) at (15.25cm, -3.75cm);
				\coordinate (point6) at (15.25cm, -16.5cm);
				\draw [arrow] (begin) -- (in1);
				\draw [arrow] (in1) -- (init);
				\draw [arrow] (init) -- (eval);
				\draw [arrow] (eval) -- (decidewhile);
				\draw [arrow] (decidewhile) -- node [left,xshift=1.5em,yshift=1em] {Yes} (popreduction);
				\draw [arrow] (popreduction) |- (op1);
				\draw [arrow] (popreduction) -- (op2);
				\draw [arrow] (popreduction) |- (op3);
				\draw [arrow] (op1) -| (updatepop);
				\draw [arrow] (op2) -- (updatepop);
				\draw [arrow] (op3) -| (updatepop);
				\draw (updatepop) -- (point1);
				\draw [arrow] (point1) |- (updatemember);
				\draw [arrow] (updatemember) -- (sort);
				\draw [arrow] (sort) -- (decidels);
				\draw [arrow] (decidels) -- node [right,xshift=0em,yshift=0em] {Yes}(localsearch);
				\draw [arrow] (localsearch) -- (comparels);
				\draw [arrow] (comparels) -- node [above,xshift=0em,yshift=0em] {Yes} (updatebest);
				\draw [arrow] (decidels) -| node [above,xshift=-8em,yshift=0em] {No} (decidewhile);
				\draw (comparels) -- (point3);
				\draw [arrow] (updatebest) -| (decidewhile);
				\draw (point3) -- node [above,xshift=4.5em,yshift=0em] {No}(updatels);
				\draw (updatels) -| (point4);
				\draw (decidewhile) -- node [above,xshift=0em,yshift=0.125em] {No}(point5);
				\draw (point5) -- (point6);
				\draw [arrow] (point6) -- (out);
				\draw [arrow] (out) -- (end);
				
		\end{tikzpicture}}
		\caption{Flowchart of IMODE algorithm \cite{imode,mendoza2022adjusting}.}
		\label{fig:imode_flowchart}
	\end{center}
\end{figure*}
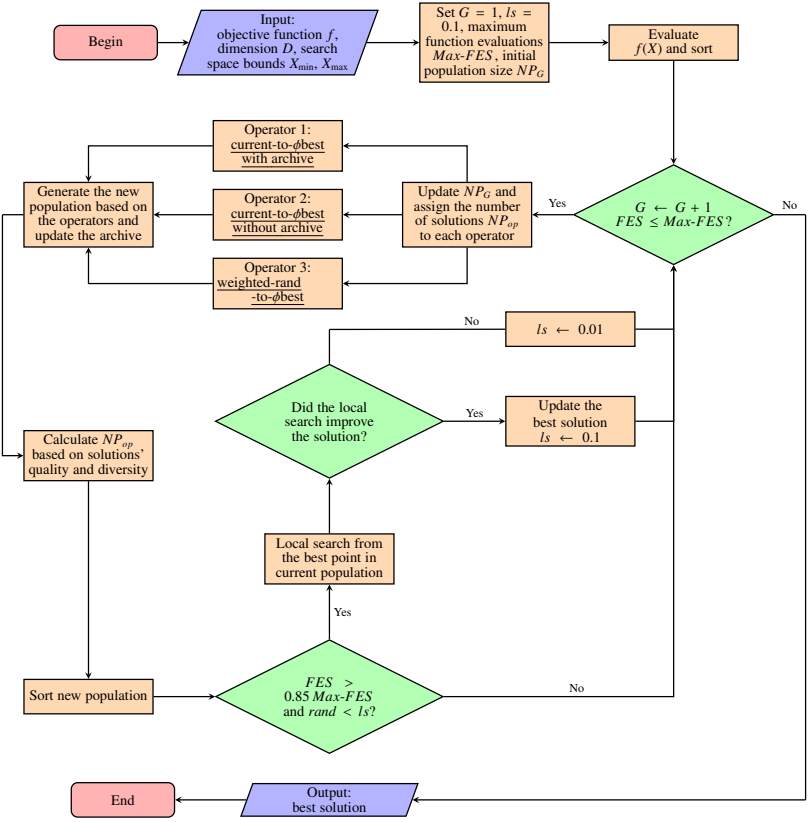

The Improved Multi-Operator Differential Evolution (IMODE) algorithm \cite{imode} is an enhanced optimization algorithm that uses the benefits of multiple differential evolution operators, with more emphasis placed on the best-performing operator. Based on the quality of solutions and the diversity of corresponding subpopulations, the superior operator is selected by using adaptively differing population sizes. IMODE also uses sequential quadratic programming (SQL) as a local search in the later stages of the evolutionary process. As an additional optimization strategy, IMODE uses adaptation mechanisms to determine parameter values and randomly chooses between binomial and exponential crossover. In 2020, IMODE ranked first in the CEC 2020 Competition on Single Objective Bound Constrained Numerical Optimization, proving its success as an optimization algorithm \cite{imode}.

The inputs of IMODE are the objective function, dimension, and bound constraints. At the beginning of the IMODE process, an initial population from the search space is generated. The objective function using each population member is calculated and then sorted in ascending order. IMODE divides the population into several sub-populations, which are all evolved using three mutation operators. Archiving is done to preserve population diversity. {In particular, inferior solutions replaced during the evolution process are stored in the archive, which serves as an external memory. These archived individuals are later used in the mutation step to introduce new variations and prevent premature convergence. This mechanism helps the algorithm maintain exploration capability and improves its ability to escape local minima. }
The three operators are given as follows:

\begin{itemize}
    \item Operator 1 ({\textit{current-to-best with archive}}): $$v_{i, j}=x_{i, j}+F_i \times\left(x_{\phi, j}-x_{i, j}+x_{r_1, j}-x_{r_2, j}\right),$$
    \item Operator 2 ({\textit{current-to-best without archive}}): $$ v_{i, j}=x_{i, j}+F_i \times\left(x_{\phi, j}-x_{i, j}+x_{r_1, j}-x_{r_3, j}\right),$$
    \item Operator 3 ({\textit{weighted-rand-to-best}}): $$v_{i, j}=F_i \times\left(x_{r_1, j}+\left(x_{\phi, j}-x_{r_3, j}\right)\right),$$
\end{itemize}
where $r_1, r_2, r_3 \neq i$ are randomly generated integers, $\vec{x}_{r_1}, \vec{x}_{r_3}$ are randomly chosen from the population, $\vec{x}_\phi$ is selected from the top $10 \%$ members of the population, and $\vec{x}_{r_2}$ is randomly selected from the union of the archive and the population. {The difference vectors $x_{r_1,j}-x_{r_2,j}$ and $x_{r_1,j}-x_{r_3,j}$ introduce stochastic perturbations that drive exploration in the population. 
The term $x_{r_1,j}-x_{r_2,j}$, where one of the indices may come from the archive, allows larger and more diverse movements that help the algorithm escape local minima and maintain population diversity. 
In contrast, $x_{r_1,j}-x_{r_3,j}$ uses only individuals from the current population, resulting in smaller, more correlated steps that promote exploitation and fine-tuning around promising regions.}

The scaling factor $F_i$ is assigned based on success-historical memory. Operators 1 and 2 move the member in the current population to the best points with and without archiving, respectively. Operator 3 is a weighted random-member-to-best operator. The size of each sub-population $\left(N P_{o p}\right.$, op $\left.=1,2,3\right)$ is iteratively adjusted based on the diversity of the sub-populations and the quality of the solutions. The size of the population per generation $\left(N P_G\right)$ is also reduced linearly. After the mutation operators are carried out, crossover is implemented randomly to create a new set of solutions.

To speed up the convergence of IMODE, a local search is implemented when the number of function evaluations exceeds $85 \%$ of the maximum function evaluations. Figure \ref{fig:imode_flowchart} shows the flowchart of the IMODE algorithm. For a detailed discussion of IMODE, we refer the readers to \cite{imode}.
\begin{figure}[!ht]
    \centering
    \includegraphics[width=0.485\textwidth]{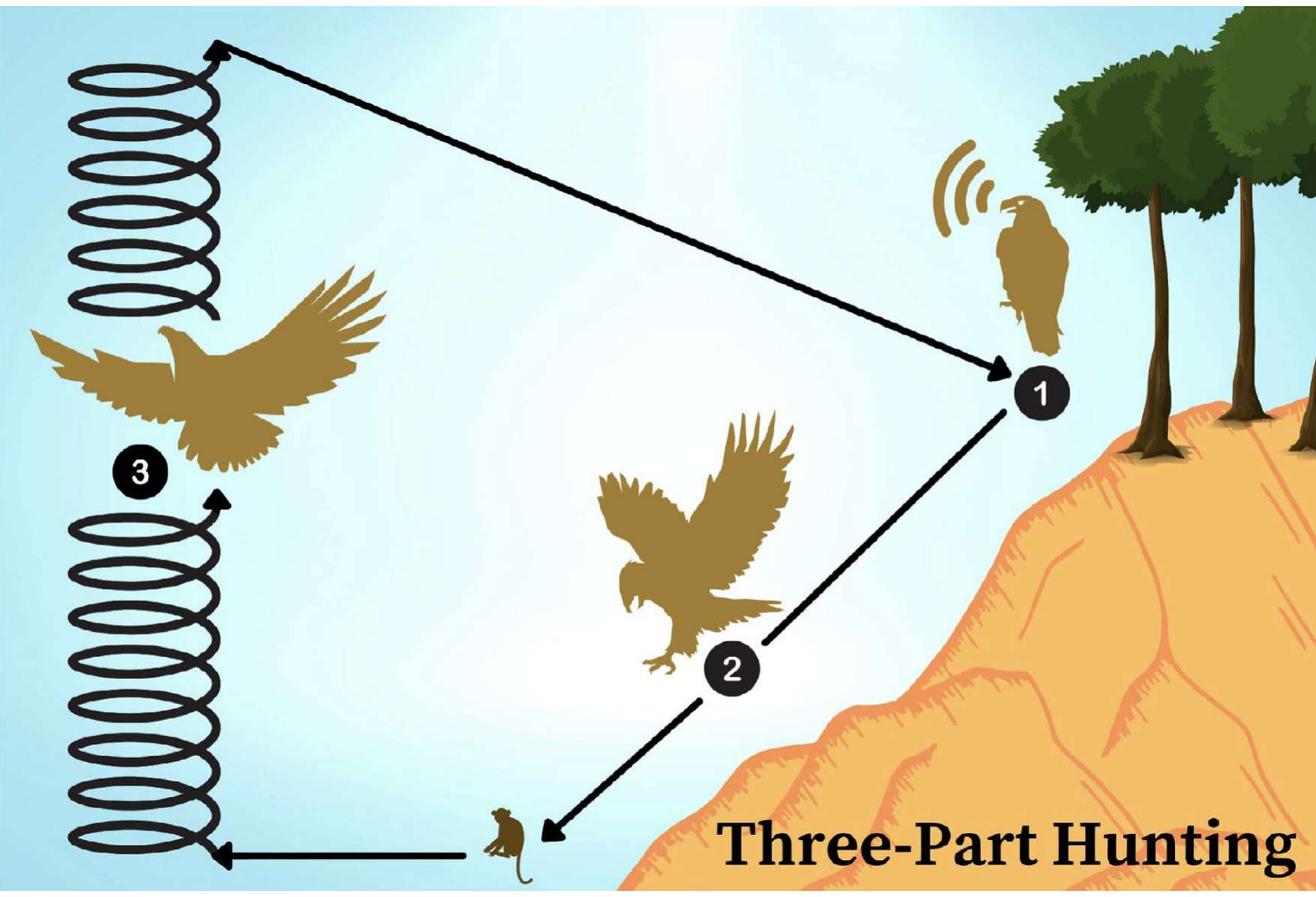}
    \caption{The three-stage hunting behavior of the Philippine Eagle begins with a preparatory phase, during which it perches and vocalizes. This is followed by the hunting action itself, as the eagle dives from its perch. Finally, it ascends in a circular motion to return to its original position \cite{bhlpart209359}.}
    \label{fig:PhEagleHunting}
\end{figure}

We take the objective function for IMODE to be the likelihoods given by Eqs. (\ref{eq:loglike_cc}-\ref{eq:loglike_rsd}). We consider an initial population size of 54, minimum number of individuals 4, maximum number of function evaluations (Max-FES) 500, and local search probability 0.1. Max-FES fixes the minimum and maximum local search evaluations \cite{imode}. For more details, our implementation can be found in \href{https://github.com/ErikaAntonette/Philippine-Eagle-Optimization-Algorithm}{github.com/ErikaAntonette/Philippine-Eagle-Optimization-Algorithm}.

{We use the default parameter values provided in the main reference paper of IMODE \cite{imode}, which are listed in Table \ref{tab:IMODE}.}
\begin{table}[ht]
\captionof{table}{{The default values of the hyperparameters of IMODE that are used in the simulations.}}
\centering\resizebox{.45\textwidth}{!}{%
\begin{tabular}{@{}cc@{}}
\toprule
Parameter                              & Value              \\ \midrule
dimension                              &          3          \\
maximum number of function evaluations &            500        \\
initial population size                        &          54         \\
minimum population size 		&          4         \\
minimum number of function evaluations to begin local search & 425 \\
local search probability                              & 0.1 \\
archive rate                                   & 2.6         \\
memory size                                  & 60         \\ \bottomrule
\end{tabular}}
\label{tab:IMODE}
\end{table}

\subsection{Philippine Eagle Optimization Algorithm}
\label{subsec:pea}

\begin{figure*}[!ht]

\tikzstyle{decision} = [diamond, draw, fill=blue!20, text width=13em, text centered, inner sep=4pt, fill=green!30]
\tikzstyle{cloud} = [rectangle, draw, fill=blue!20, text width=5em, text centered, rounded corners=12pt, minimum height=2em]
\tikzstyle{line} = [draw, -latex']
\tikzstyle{block} = [rectangle, draw, fill=blue!20, text width=18em, text centered, minimum height=3.5em]
\tikzstyle{put} = [trapezium, draw, text width=10.5em, fill=blue!20, minimum height=2em, trapezium left angle=50, trapezium right angle=130, text centered]
\tikzstyle{block2} = [rectangle, draw, fill=blue!20, text width=15em, text centered, minimum height=3.5em]
\tikzstyle{startstop} = [rectangle, rounded corners, minimum width=3cm, minimum height=1cm,text centered, draw=black, fill=red!30]
\tikzstyle{io} = [trapezium, trapezium left angle=70, trapezium right angle=110, minimum width=0.5cm, minimum height=2cm, text centered, draw=black, fill=blue!30,  text width=11em]
\tikzstyle{process} = [rectangle, minimum width=3.1cm, minimum height=1cm, text centered, text width=17.8em, draw=black, fill=orange!30]
\tikzstyle{process2} = [rectangle, minimum width=3cm, minimum height=1cm, text centered, text width=15.8em, draw=black, fill=orange!30]
\tikzstyle{decision2} = [diamond, minimum width=3cm, minimum height=1cm, text centered, draw=black, fill=green!30]
\tikzstyle{arrow} = [thick,->,>=stealth]

\begin{center}
\resizebox{0.95\textwidth}{!}{
\begin{tikzpicture}[thick, every node/.style={scale=0.75},node distance = 2.5cm, auto]
    \node [startstop] (begin) {Begin};
	\node [io, right of=begin,node distance=5.5cm] (in1) {Input objective function $f$, dimension $D$, search space bounds $X_{\textrm{min}}$, $X_{\textrm{max}}$};
    \node [process, right of=in1,node distance=7.75cm] (init) {Initialize maximum function evaluations $N_\textrm{max}$, initial eagle population size $S_0$, local food size $S_{\textrm{loc}}$};
	\node [process, right of=init,node distance=8.25cm] (set) {Set equal probabilities of the three operators \emph{Movement, Mutation I, and Mutation II} and initialize memory elements for the scaling factors $F$};
    \node [process, below of=set,node distance=2.8cm] (gen1) {Generate initial population of eagles $X$ and select the best eagle $X^\star$ of the population based on function value};
    \node [process, left of=gen1,node distance=9cm] (conduct) {Conduct local search for the best food $Y^\star$ of the {best~eagle} using interior point method};
 \node [process, below of=conduct] (update1) {Update eagle population size $S$ using linear population size reduction};
    \node [process, below of=update1] (divide) {Divide eagle population into three subpopulations using the probabilities of the three operators};

    \node [process2, below of=divide, xshift=-17em,yshift=-2em] (op1) {Apply \emph{Movement} operator to the first subpopulation of eagles:\\ $(X_{\textrm{new}})_i =  X_i  +  F_i \cdot (X^\star - X_i \,+\, X_{r_1} - X_{\textrm{arc}}\, +  e^{-d^2} \cdot (X_{\textrm{near}} - X_i) )$};
\node [process2, below of=divide,yshift=1.6em,node distance=4cm] (op2) {Apply \emph{Mutation I} operator to the second subpopulation of eagles: \\ $(X_{\textrm{new}})_i =  F_i \cdot (X_{r_1} +  {X^\star - X_{r_2})}  + S \cdot L(D)$};
\node [process2, below of=divide, xshift=17em,yshift=-2em] (op3) {Apply \emph{Mutation II} operator to the third subpopulation of eagles:\\ $(X_{\textrm{new}})_i  =  F_i \cdot (\hat{X} + X^\star - X_{\textrm{mean}})$};

    \node [process, below of=op2,node distance=3.7cm] (gen2) {Generate new population of eagles by replacing old eagles with the new eagles $X_\textrm{new}$ that gained improvement in function value};
    \node [process, below of=gen2,node distance=3cm] (conduct2) {Update the best eagle $X^\star$ of the population and search for the new best food $Y^\star$ of the updated best eagle using interior point method};
    \node [decision, below of=conduct2, node distance=5.4cm] (decide) {Is the function value error $|f(Y^\star)-f_{\textrm{true}}|$ less than $10^{-8}$?};
   \node [decision, below of=decide,  node distance=8.5cm] (decide2) {Is the current number of function evaluations  $N$ greater that the maximum function evaluations?};
\node [io, left of=decide, node distance=7cm,yshift=-10em,xshift=3.5em] (in2) {Output~searched~food~$Y^\star$ of the best eagle and its function value $f(Y^\star)$};
\node [startstop, left of=in2,xshift=-8em] (end) {End};
    \node [process, right of=decide, node distance=15cm, xshift=-10.2em] (update3) {Update the probabilities of the three operators \emph{Movement, Mutation I, and Mutation II} based on the improvement rate of each operator};
    \node [process, above of=update3, node distance=15cm, yshift=-25em] (update4) {Update the memory elements for the scaling factors using the weighted Lehmer mean of the scaling factors of improved eagles};
    \draw [arrow] (begin) -- (in1);
    \draw [arrow] (in1) -- (init);
    \draw [arrow] (init) -- (set);
    \draw [arrow] (set) -- (gen1);
    \draw [arrow] (gen1) -- (conduct);
    \draw [arrow] (conduct) -- (update1);
    \draw [arrow] (update1) -- (divide);
    \draw [arrow] (divide) -| (op1);   
    \draw [arrow] (divide) -| (op3);   
    \draw [arrow] (divide) -- (op2);   
\draw [arrow] (op2) -- (gen2);
\draw [arrow] (op1) |- (gen2);
\draw [arrow] (op3) |- (gen2);
    \draw [arrow] (gen2) -- (conduct2);
    \draw [arrow] (conduct2) -- (decide);
 \draw [arrow] (decide) -- (decide2);
 \draw [arrow] (in2) -- (end);
    \draw [arrow] (decide) -| node [above,xshift=3em] {Yes}(in2);
 \draw [arrow] (decide) -- node [right,xshift=0.3em] {No}(decide2);
	\draw [arrow] (decide2) -| node [above,  xshift=3em] {Yes}(in2);
    \draw [arrow] (decide2) -| node [above,  xshift=-15.5em] {No}(update3);
     \draw [arrow] (update3) -- (update4);
    \draw [arrow] (update4) |-(update1);
\end{tikzpicture}}
		\caption{Flowchart of PEOA \cite{Enriquez_Mendoza_Velasco_2022}.}
		\label{fig:peoa_flowchart}
\end{center}

\end{figure*}
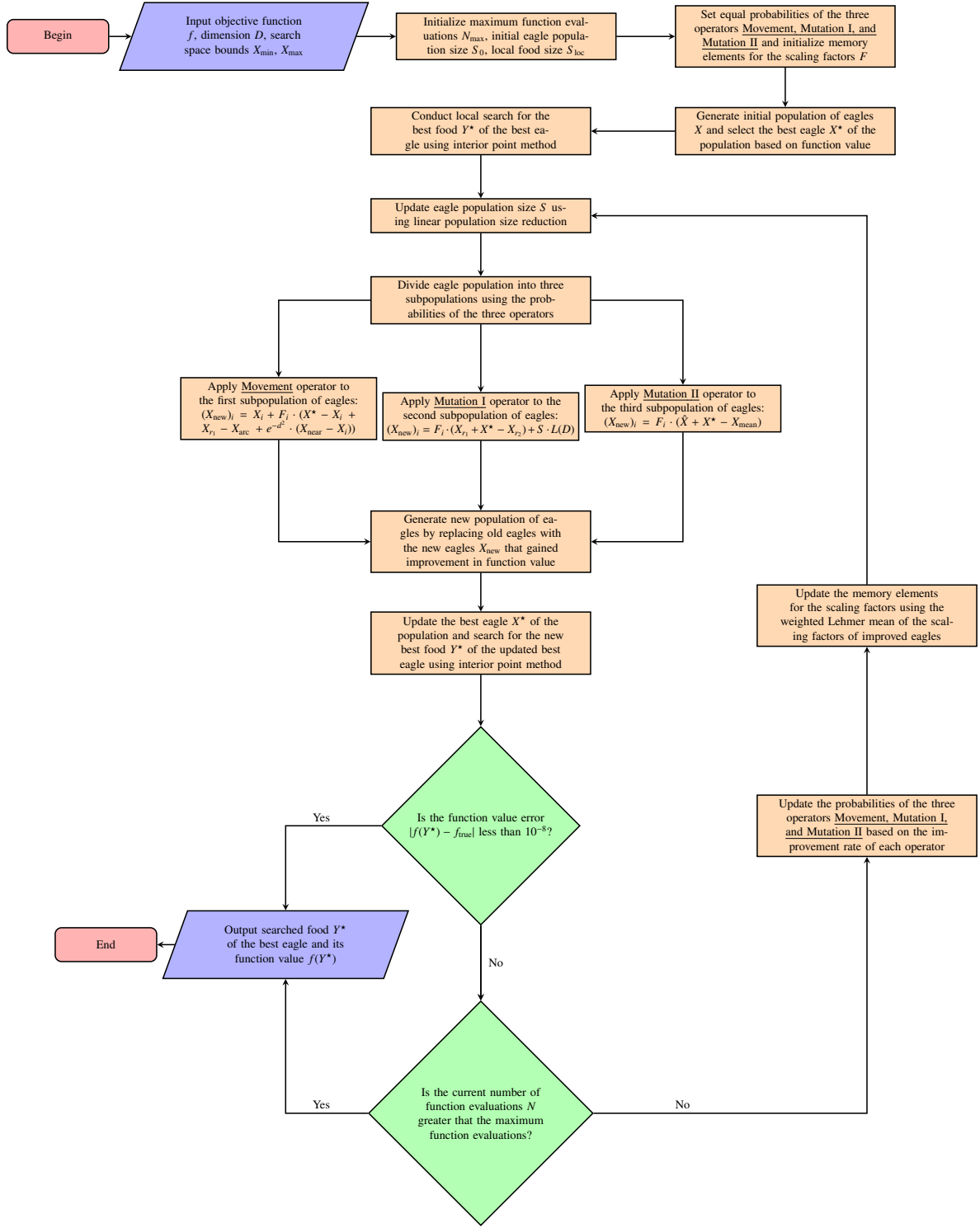

The Philippine Eagle Optimization Algorithm (PEOA) \cite{Enriquez_Mendoza_Velasco_2022} is a novel, meta-heuristic, and population-based search algorithm inspired by the territorial hunting behavior of the Philippine Eagle (Figure \ref{fig:PhEagleHunting}). In \cite{Enriquez_Mendoza_Velasco_2022}, twenty optimization test functions of varying properties on modality, separability, and dimension were solved using PEOA, and the results were compared to those
obtained by thirteen other optimization algorithms including GA and IMODE. This benchmark has shown PEOA outperforming a wide variety of established algorithms in terms of accuracy and precision in finding the optimal solution of tested functions. In addition, PEOA uses the least number of function evaluations compared to the other algorithms, indicating that it employs a computationally efficient process.

The strategy of PEOA is as follows (Figure \ref{fig:peoa_flowchart}). Starting with an initial random population of eagles in a given search space, the best eagle is selected and undergoes a local food search using the interior point method as its means of exploitation. The population is then divided into three subpopulations, and each subpopulation is assigned an operator which aids in the exploration. Once the respective operators are applied, the new eagles with improved function values replace the older ones. The best eagle of the population is then updated and conducts a local food search again. These steps are done iteratively, and the food searched by the final best eagle is the optimal solution of the search space. Thus, PEOA  uses three different global operators for its exploration strategy, and also undergoes intensive local search every iteration, contributing to its strong exploitation ability.

{Because PEOA is inspired by the Philippine Eagle, several terms in the algorithm are derived from this analogy. For example, an ``eagle" is a candidate solution to the given optimization problem, similar to a ``chromosome" in GA. The eagle with the least function value (in a minimization problem) is called the ``best eagle" of the eagle population. The best eagle then conducts a local search, which is the use of an optimizer (\textit{e.g.}, the interior point method) to find the minimum point in a small region containing the best solution. In PEOA, the minimum point found by the best eagle during its local search is its ``food," and consequently, the local search is also called the ``local food search." 
If the stopping criterion has not yet been satisfied, PEOA proceeds to generate a new eagle population using three operators, namely ``Movement," ``Mutation I," and ``Mutation II" (see Figure \ref{fig:peoa_flowchart}). These operators work in a similar manner as the crossover, mutation, and selection operators of GA.}

As with GA and IMODE, we take the PEOA objective function to be the likelihoods (\ref{eq:loglike_cc}-\ref{eq:loglike_rsd}) and generate the uncertainty by applying the method iteratively to bootstrapped data. We use default configurations with PEOA, with a minimum eagle population size of 5, cluster size factor 0.04, and Max-FES 500 \cite{Enriquez_Mendoza_Velasco_2022}. Our detailed implementation of PEOA can be found in \href{https://github.com/ErikaAntonette/Philippine-Eagle-Optimization-Algorithm}{github.com/ErikaAntonette/Philippine-Eagle-Optimization-Algorithm}. {Lastly, we use the default parameter values of the PEOA adopted in its main reference paper \cite{Enriquez_Mendoza_Velasco_2022}, which are listed in Table \ref{tab:PEOA}.}
\begin{table}[ht]
\captionof{table}{{The default values of the hyperparameters of PEOA that are used in the simulations.}}
\centering\resizebox{.45\textwidth}{!}{%
\begin{tabular}{@{}cc@{}}
\toprule
Parameter & Value \\ 
\midrule
dimension & 3 \\ 
maximum number of function evaluations & 500 \\
initial population size                        &          180         \\
minimum population size & 5 \\
maximum number of function evaluations during local search & 90 \\
cluster size factor & 0.04 \\ 
archive rate & 2.6 \\ 
memory size & 60 \\ 
\bottomrule
\end{tabular}}
\label{tab:PEOA}
\end{table}

\subsection{Uncertainty Analysis}
\label{subsec:uncertainty_analysis}

\begin{algorithm}[h!]
\caption{Uncertainty Quantification in Metaheuristic Optimization}
\begin{algorithmic}[1]
\renewcommand{\algorithmicrequire}{\textbf{Input:}}
\renewcommand{\algorithmicensure}{\textbf{Output:}}
\REQUIRE Mock data $ (z_{\rm mock}, y_{\rm mock}) $, standard deviation $ \Delta y_{\rm mock} $, number of realizations $ N$
\ENSURE Distribution of optimizers
\STATE Initialize set of perturbed data sets $ D = \emptyset $ and set of optimizers $ \mathcal{O} = \emptyset $
\FOR{$ i = 1 $ to $ N $}
    \STATE Generate perturbed data: $ y_{\rm mock}^{(i)} \sim \mathcal{N}(y_{\rm mock}, \Delta y_{\rm mock}) $
    \STATE Store $ (z_{\rm mock}, y_{\rm mock}^{(i)}) $ in $ D $
\ENDFOR
\FOR{$ i = 1 $ to $ N $}
    \STATE Optimize fitness functions in Eqs. (\ref{eq:loglike_cc}--\ref{eq:loglike_rsd}) using a metaheuristic algorithm
    \STATE Store resulting optimizer in $ \mathcal{O} $
\ENDFOR
\STATE Analyze statistical distribution of $\mathcal{O} $
\RETURN Distribution of optimizers
\end{algorithmic} 
\label{algo:uq_metaheuristic}
\end{algorithm}

Incorporating uncertainty quantification into the optimization framework of a metaheuristic algorithm allows for analysis of the statistical behavior of the estimated parameters. Given the mock data $ (z_{\rm mock}, y_{\rm mock}) $, we introduce normally distributed noise to $ y_{\rm mock} $ with  $ \Delta y_{\rm mock} $ as the standard deviation, effectively generating perturbed realizations of the observed data. This process is repeated ${\cal O}\left(10^{3-4}\right)$ times, creating a comprehensive set of data with uncertainty. For each of these data sets, we perform optimization using GA, IMODE, and PEOA on the fitness functions described in Eqs. (\ref{eq:loglike_cc}--\ref{eq:loglike_rsd}). The resulting optimizers from each iteration are stored and analyzed to characterize their statistical distributions, providing assessment into the reliability and variability of the obtained parameters under uncertainty. This is summarized in Algorithm \ref{algo:uq_metaheuristic}.

{We emphasize that the uncertainty obtained using the above described bootstrap mechanism is conceptually different compared to ones estimated by a Bayesian method like MCMC. The bootstrap approximates frequentist confidence intervals, which quantify parameter variations under repeated sampling of the data. In contrast, Bayesian intervals represent regions of parameter space with a fixed posterior probability, given the observed data and prior information.}

\section{Results}
\label{sec:results}

This section presents the main results of our work. In Section~\ref{subsec:results_CC_RSD}, we discuss the results obtained when only expansion rate and growth rate data are used. Section~\ref{subsec:results_All} extends the analysis by incorporating mock supernova measurements into the data set.

\subsection{Expansion Rate + Growth Rate}
\label{subsec:results_CC_RSD}

We examine the posterior estimates obtained when considering only expansion and growth rate observations. For each of the 500 mock data sets analyzed, posteriors were derived using MCMC, GA, IMODE, and PEOA. {The results were computed using a single machine to guarantee computational budget fairness across methods.}

\begin{figure}[h!]
    \centering
    {\includegraphics[width=0.485\textwidth]{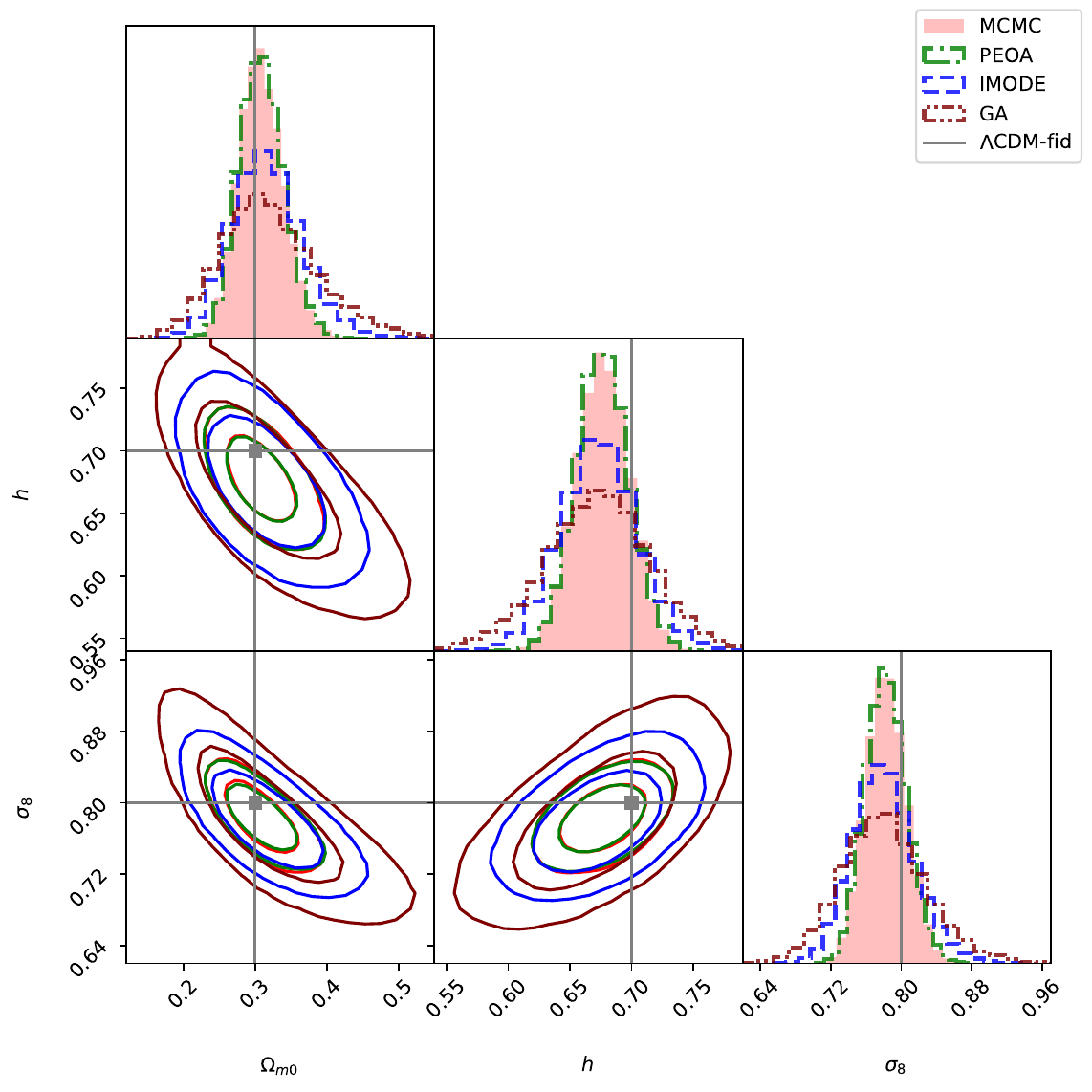}}
    \caption{Posterior estimates for a given mock data set (expansion rate + growth rate) obtained with MCMC, GA, IMODE, and PEOA.}
    \label{fig:posts_and_rec_CC_RSD_i}
\end{figure}

Figure \ref{fig:posts_and_rec_CC_RSD_i} illustrates a representative case of parameter estimates using a single mock data set (one realization).

The results demonstrate that the estimates obtained with PEOA closely match those from MCMC in both shape and precision. In terms of increasing precision, GA is the least precise, followed by IMODE, while PEOA and MCMC exhibit comparable precision. This precision hierarchy is reflected in the width of the contours or error bars. The reconstructed cosmological functions from PEOA are visually indistinguishable from those derived using MCMC. Conversely, reconstructions using GA and IMODE exhibit broader error bars, attributable to wider parameter covariances.

However, the performance varies across realizations, as seen in Figure \ref{fig:mock_data_i}. Evaluating precision and accuracy using a single mock data set does not account for variance across realizations. To address this, we quantify accuracy and precision for each method over all mock data sets. These metrics are analyzed using histograms or tabulated results to capture their statistical distributions. Additionally, we evaluate the reconstruction quality of parametric reconstructions based on the parameters covariance obtained for each method.

\begin{figure}[h!]
    \centering
    {\includegraphics[width=0.485\textwidth]{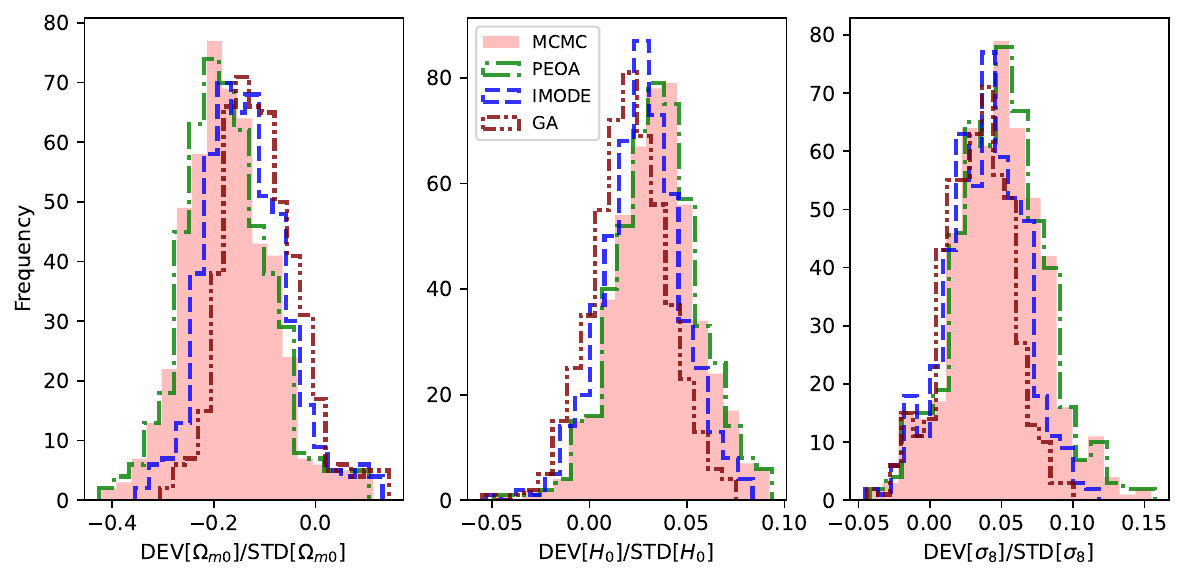}}
    {\includegraphics[width=0.485\textwidth]{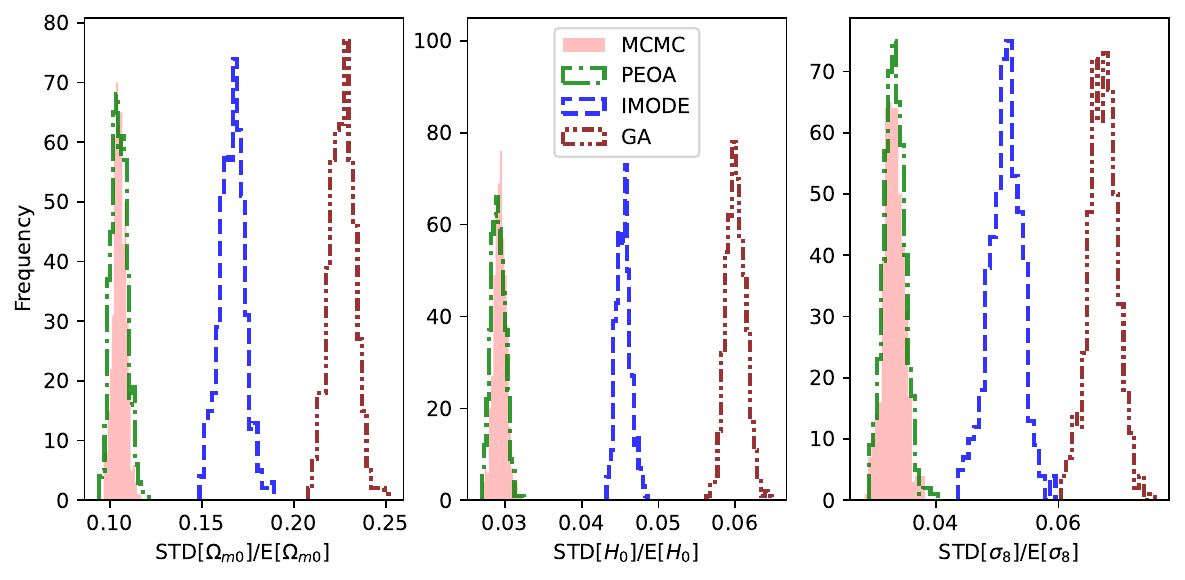}}
    \caption{Histograms for [top] accuracy, $(\overline{a} - a^{\rm true})/a^{\rm true}$, and [bottom] precision, $\Delta a/\overline{a}$, per parameter $a=(\Omega_{m0}, H_0, \sigma_8)$ for MCMC, GA, IMODE, and PEOA. $\overline{a}$ and $\Delta a^2$ denote the sample mean and variance.}
    \label{fig:accuracy_and_precision_CC_RSD}
\end{figure}

Accuracy is quantified using the relative deviation (RDEV) of each parameter $a=(\Omega_{m0}, H_0, \sigma_8)$:
\begin{equation}
    \label{eq:dev_rel}
    {\rm RDEV}[a]=\dfrac{\overline{a} - a^{\rm true}}{a^{\rm true}} \,,
\end{equation}
where $\overline{a}$ represents the sample mean and $a^{\rm true}$ the true input parameters, i.e., $\Omega_{m0}=0.3$, $H_0=70$ km s$^{-1}$Mpc$^{-1}$, and $\sigma_{8}=0.8$. {When RDEV = 0, then the sample mean lies exactly on the true values. When RDEV is positive (negative), the sample mean lies to the right (left) of the true values.} Precision is defined as:
\begin{equation}
    \label{eq:prec}
    {\rm PREC}[a]= \dfrac{{\rm STD}[a]}{{\rm AVE}[a]} =\dfrac{\Delta a}{\overline{a}} \,,
\end{equation}
measuring normalized uncertainty. {This purely positive quantity measures the width of a sample relative to its mean value. In general, small RDEV and PREC correspond to accurate and precise estimates, respectively. The question of how small or large depends on the sample and the objectives of the work. In our case we find that these two metrics are sufficiently able to distinguish the methods under study. Of course, alternative metrics are available, but we shall leave this for future studies.} Figure \ref{fig:accuracy_and_precision_CC_RSD} shows the accuracy and precision for 500 realizations.

The results reveal that $\Omega_{m0}$ estimates typically fall below the true value, reflected in negative deviations in Figure \ref{fig:accuracy_and_precision_CC_RSD}. Meanwhile, $H_0$ and $\sigma_8$ estimates generally exceed the true values. Despite these trends, all methods achieve similar accuracy, as evidenced by their comparable means across realizations. This consistency explains the similarity in histogram distributions.

Precision, on the other hand, differs across methods, as shown in the bottom panel of Figure \ref{fig:accuracy_and_precision_CC_RSD}. GA is the least precise, followed by IMODE, while PEOA and MCMC demonstrate higher precision. This distinction aligns with Figure \ref{fig:posts_and_rec_CC_RSD_i}, where the precision order can be visually inferred from the contour sizes. For $\Omega_{m0}$, average precision values are approximately 11\%, 23\%, 17\%, and 11\% for MCMC, GA, IMODE, and PEOA, respectively. Similar trends are observed for $H_0$ and $\sigma_8$, highlighting consistent precision hierarchies across parameters and covariances. {Note that the bottom panel of Figure \ref{fig:accuracy_and_precision_CC_RSD} (PREC histograms) additionally attests to the stability of the bootstrap interval across realizations; in the sense that the histograms of the EAs can be visualized within compact distribution functions with well defined moments.}

To evaluate reconstruction quality, we employ a chi-squared-like distance metric:
\begin{equation}
\label{eq:goodness_rec}
    D_0(f(Z)|g(Z))= \dfrac{1}{N_Z}\sum_Z \left( \dfrac{ f(Z)-g(Z) }{\sqrt{\Delta f(Z)^2 + \Delta g(Z)^2}} \right)^2 \,,
\end{equation}
previously used in \cite{Bernardo:2021cxi, Bernardo:2022pyz}. Here, $D_0$ measures the weighted distance between two functions $f$ and $g$ across grid points $Z$. Reconstructions with $D_0 \sim 1/2$ are considered optimal, while $D_0 \ll 1/2$ or $D_0 \gg 1/2$ suggest over-fitting or under-fitting, respectively.

\begin{figure}[h!]
    \centering
    \includegraphics[width=0.485\textwidth]{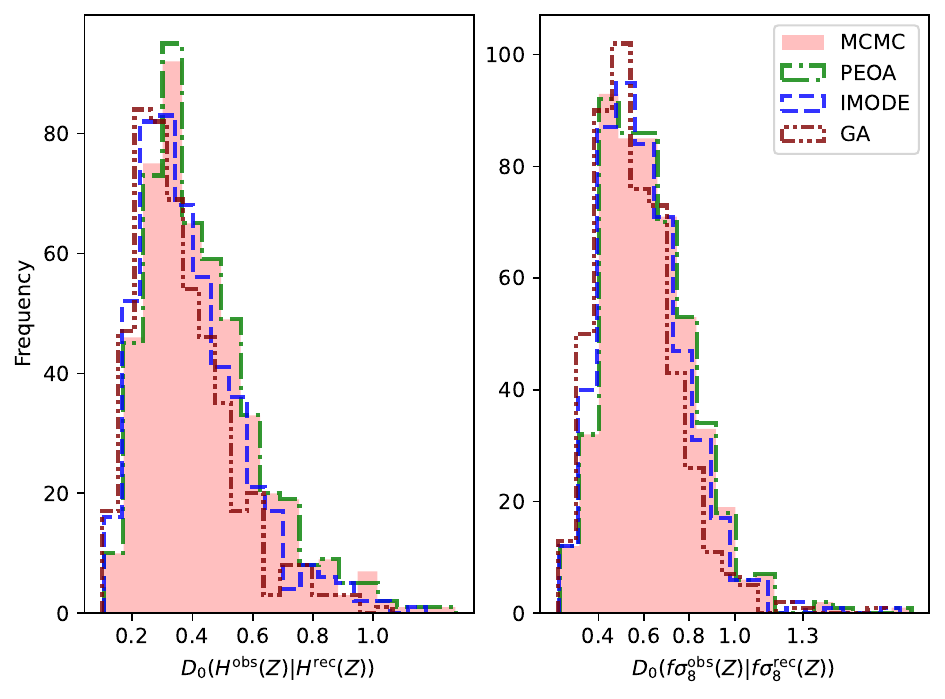}
    \caption{Histograms for the goodness \eqref{eq:goodness_rec} of the reconstructed cosmological functions: the expansion rate and the growth rate.}
    \label{fig:goodness_of_recs_CC_RSD}
\end{figure}

Figure \ref{fig:goodness_of_recs_CC_RSD} shows $D_0$ values across 500 realizations, indicating that all methods yield reconstructions consistent with the ideal case. Note that the cosmological reconstructions are obtained by resampling the parameters using the measured covariances for each mock data set. The agreement stems from parameter resampling using measured covariances for each mock data set, affirming the robustness of the methods.

In summary, while MCMC, GA, IMODE, and PEOA exhibit similar accuracy in parameter estimation and reconstruction, their precision varies significantly. GA is the least precise, followed by IMODE, whereas PEOA and MCMC demonstrate comparable, superior precision.

\subsection{Expansion Rate + Growth Rate + Supernovae}
\label{subsec:results_All}

\begin{figure}[h!]
    \centering
    {\includegraphics[width=0.485\textwidth]{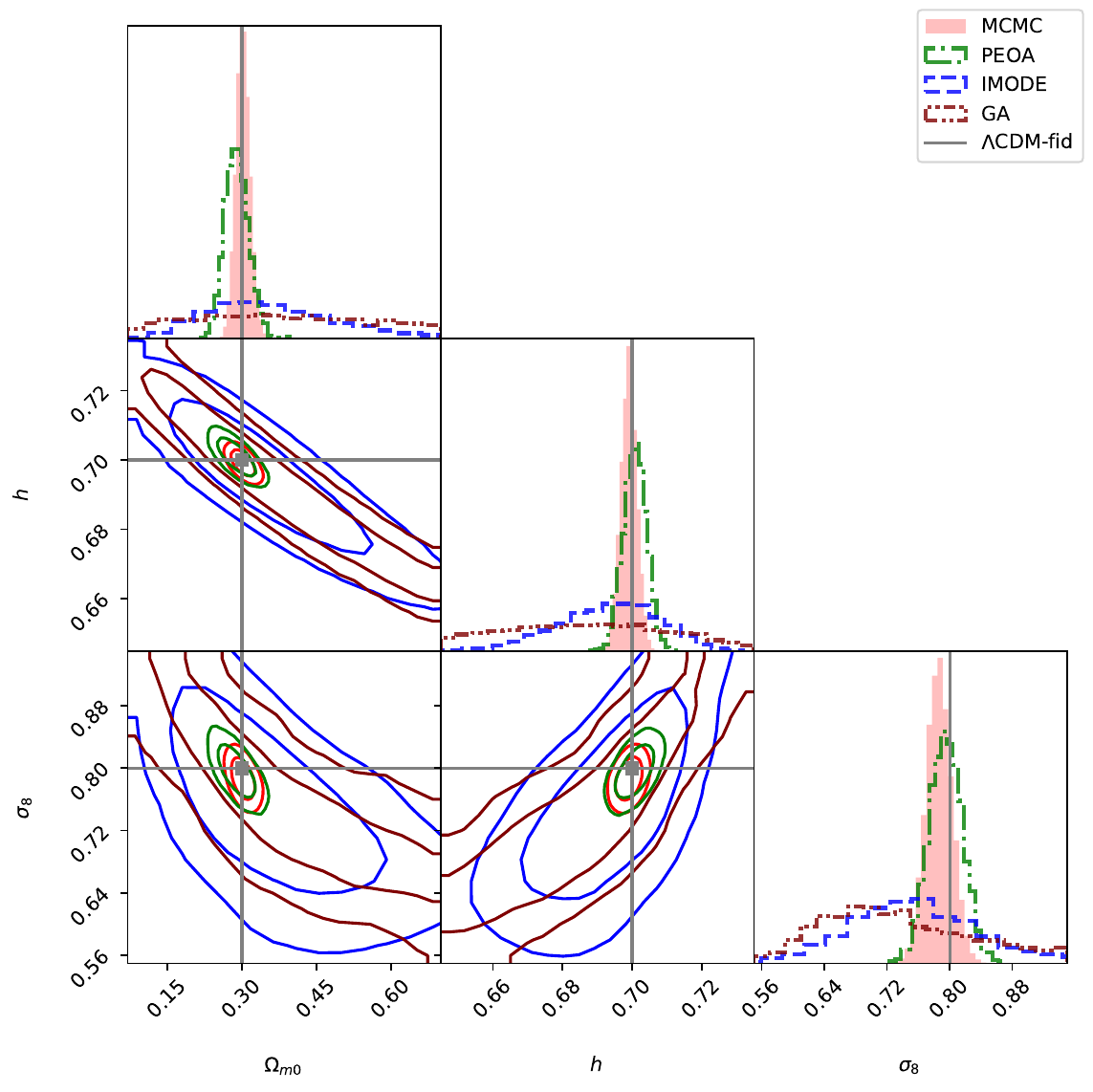}}
    \caption{Posterior estimates for a given mock data (expansion rate + growth rate + supernovae) obtained with MCMC, GA, IMODE, and PEOA.}
    \label{fig:posts_and_rec_CC_RSD_SNe_i}
\end{figure}

In this section, we incorporate mock SNe measurements (see Section \ref{sec:mock_data}) in addition to expansion rate and growth rate data to evaluate the robustness of the analysis. The typical posterior constraints (one realization/mock data) for this data combination are illustrated in Figure \ref{fig:posts_and_rec_CC_RSD_SNe_i}.

It is worth emphasizing that the parameters used here are identical to those in Section \ref{subsec:results_CC_RSD}. For MCMC, the inclusion of supernovae data improves both the accuracy and precision of parameter constraints, as expected from statistical considerations. PEOA exhibits a similar trend. However, counterintuitively, the constraints obtained with IMODE and GA degrade in precision despite the addition of more stringent data. This can be attributed to the rate of convergence of EAs towards a true solution, which is sensitive to the operators used. PEOA, by design, outperforms IMODE and GA in this aspect. Further details will be discussed in Section \ref{sec:discussion}.

\begin{figure}[h!]
    \centering
    {\includegraphics[width=0.485\textwidth]{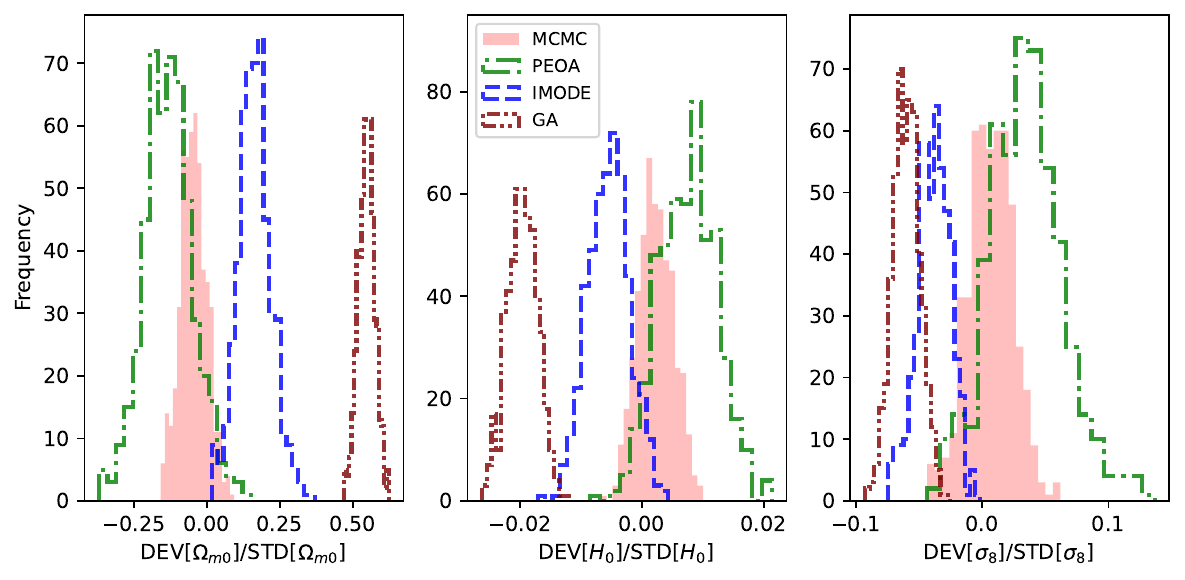}}
    {\includegraphics[width=0.485\textwidth]{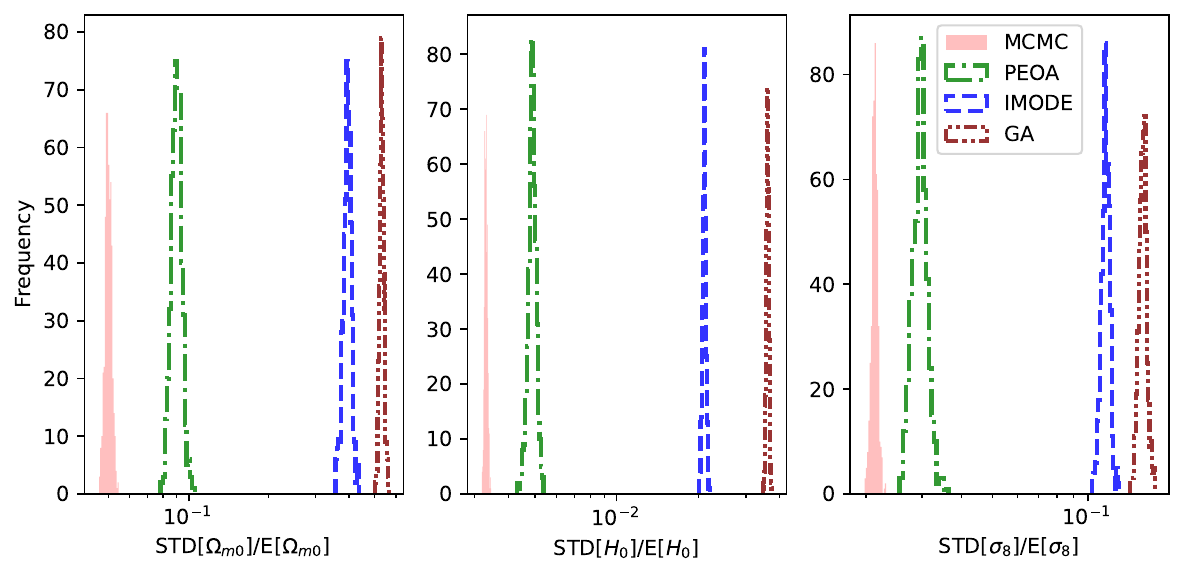}}
    \caption{Histograms for multiple mock data realizations of the [top] accuracy ($(\overline{a} - a^{\rm true})/a^{\rm true}$) and [bottom] precision ($\Delta a/\overline{a}$) per parameter $a=(\Omega_{m0}, H_0, \sigma_8)$ for each of the methods (MCMC, GA, IMODE, PEOA).}
    \label{fig:accuracy_and_precision_CC_RSD_SNe}
\end{figure}

When multiple realizations are analyzed, the impact of SNe data becomes clear. For MCMC, the precision of $\Omega_{m0}$ improves from $11\%$ to $5\%$, $H_0$ from $3\%$ to $0.3\%$, and $\sigma_8$ from $3\%$ to $2\%$. For PEOA, similar improvements are observed, with the precision of $\Omega_{m0}$ improving from $11\%$ to $9\%$, $H_0$ from $3\%$ to $0.5\%$, and $\sigma_8$ remaining at $3\%$. The enhancement in $H_0$ precision highlights the reliability of SNe data in constraining cosmological models. In contrast, IMODE and GA experience significant drops in precision, with IMODE’s precision for $\Omega_{m0}$ worsening from $17\%$ to $39\%$, $H_0$ improving slightly from $5\%$ to $2\%$, and $\sigma_8$ worsening from $5\%$ to $11\%$. For GA, the precision of $\Omega_{m0}$ degrades from $23\%$ to $53\%$, $H_0$ improves slightly from $6\%$ to $4\%$, and $\sigma_8$ worsens from $7\%$ to $15\%$. These trends are consistent across multiple mock data sets and are visualized in Figure \ref{fig:accuracy_and_precision_CC_RSD_SNe}.

Compared to the previous section, the accuracy histograms now show pronounced differences among the methods. MCMC and PEOA maintain acceptable accuracy, with their estimates centered around the true values within reasonable margins. Additionally, the inclusion of SNe improves their precision. However, IMODE and GA demonstrate a stark contrast, with reductions in both precision and accuracy, contrary to expectations. This will be explained in the next section.

\begin{figure}[h!]
    \centering
    \includegraphics[width=0.485\textwidth]{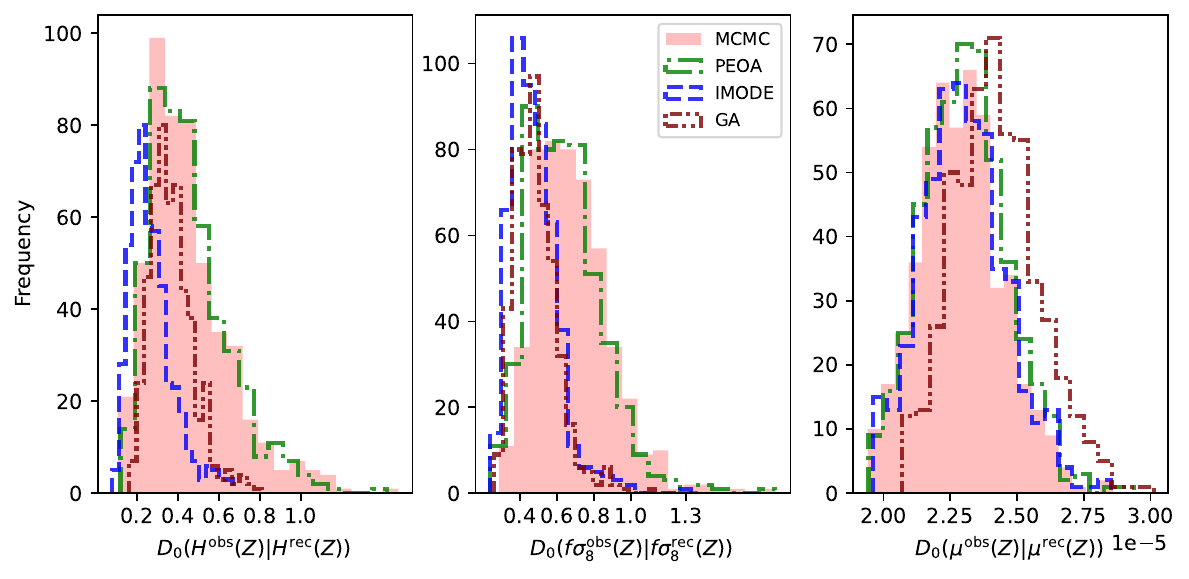}
    \caption{Histograms for the goodness \eqref{eq:goodness_rec} of the reconstructed expansion rate, growth rate, and distance modulus.}
    \label{fig:goodness_of_recs_CC_RSD_SNe}
\end{figure}

Despite the differences highlighted by accuracy and precision metrics, the goodness-of-reconstruction metric \eqref{eq:goodness_rec} (Figure \ref{fig:goodness_of_recs_CC_RSD_SNe}) shows comparable results for all methods. This metric, while valuable, does not fully capture the discrepancies between methods revealed by other metrics. IMODE and GA exhibit tendencies to overfit the expansion and growth rate data relative to MCMC and PEOA, which produce nearly indistinguishable histograms. All methods appear to overfit the SNe data, indicating the need to assess reconstructions using diverse evaluation metrics \cite{DiValentino:2025sru, H0RecOlympics}.

In summary, this section highlights the counterintuitive response of certain EAs, such as IMODE and GA, to the inclusion of more stringent data, while demonstrating the robustness of PEOA as a supporting/alternative tool to MCMC for cosmological parameter estimation and reconstruction. Further discussion on the hyperparameters of EAs will follow in the next section, providing deeper insights into their performance.

\section{Discussion}
\label{sec:discussion}

Table \ref{tab:summary_stats} summarizes the ensemble averages (over 500 mock data sets) of the accuracy and precision of MCMC, GA, IMODE, and PEOA in $\Lambda$CDM cosmological parameter estimation. Note that these statistics are the same as Figures \ref{fig:accuracy_and_precision_CC_RSD} and \ref{fig:accuracy_and_precision_CC_RSD_SNe}.

\begin{table*}[h!]
    \centering
    \caption{Ensemble averages (multiple realizations) of accuracy (Eq. \eqref{eq:dev_rel}) and precision (Eq. \eqref{eq:prec}) of MCMC, GA, IMODE, and PEOA for $\Lambda$CDM parameter estimation. $H(z)$+$f\sigma_8(z)$ refers to when only expansion rate and growth rate measurements are considered; and $H(z)$+$f\sigma_8(z)$+$\mu(z)$ when supernovae is additionally considered.
    }
    \resizebox{0.9\textwidth}{!}{%
    \begin{tabular}{|c|c|cc|cc|cc|}
    \hline
    Data set & Method & RDEV[$\Omega_{m0}$] & PREC[$\Omega_{m0}$] & RDEV[$H_0$] & PREC [$H_0$] & RDEV[$\sigma_8$] & PREC[$\sigma_8$] \\ \hline \hline
    \multirow{4}{*}{$H(z)$+$f\sigma_8(z)$} &
    MCMC & $-17\%$ & $11\%$ & $3\%$ & $3\%$ & $5\%$ & $3\%$ \\ 
    & GA & $-11\%$ & $23\%$ & $2\%$ & $6\%$& $3\%$ & $7\%$ \\ 
    & IMODE & $-14\%$ & $17\%$ & $3\%$ & $5\%$ & $4\%$ & $5\%$ \\ 
    & PEOA & $-17\%$ & $11\%$ & $3\%$ & $3\%$ & $5\%$ & $3\%$ \\ \hline
    \multirow{4}{*}{$H(z)$+$f\sigma_8(z)$+$\mu(z)$} &
    MCMC & $-5\%$ & $5\%$ & $0.2\%$ & $0.3\%$ & $0.7\%$ & $2\%$ \\ 
    & GA & $5\%$ & $53\%$ & $-2\%$ & $4\%$ & $-6\%$ & $15\%$ \\ 
    & IMODE & $16\%$ & $39\%$ & $-0.5\%$ & $2\%$ & $-4\%$ & $11\%$ \\ 
    & PEOA & $-13\%$ & $9\%$ & $0.8\%$ & $0.5\%$ & $3\%$ & $3\%$ \\ \hline
    \end{tabular}
    }
    \label{tab:summary_stats}
\end{table*}

With expansion and growth rate data ($H(z)$+$f\sigma_8(z)$), all methods performed comparably well. However, when supernovae data ($\mu(z)$) was added, PEOA consistently outperformed IMODE and GA, matching the performance of MCMC. 

This can be attributed to how PEOA was originally formulated, with certain aspects derived from IMODE and its several source algorithms \cite{Enriquez_Mendoza_Velasco_2022}. These algorithms include United Multi-Operator Evolutionary Algorithms-II (UMOEAs-II), Adaptive Differential Evolution Algorithm (JADE), and Success-History Based Adaptive Differential Evolution with Linear Population Size Reduction Algorithm (L-SHADE) \cite{imode}.

For instance, PEOA uses the Interior Point Method for its local food search, which can increase its exploitation ability, as claimed in UMOEAs-II. Moreover, the Movement Operator of PEOA, based on an operator from JADE, has good searching ability and can prevent traps in local minima. The Mutation I and II operators of PEOA feature not only characteristics of the Philippine Eagle but also strengthen the exploration ability of the algorithm. These two operators were constructed based on UMOEAs-II, the Firefly Algorithm, and the Harris Hawks Optimization Algorithm \cite{Enriquez_Mendoza_Velasco_2022}.

Therefore, PEOA can be considered a further improved version of all the aforementioned algorithms, resulting in its superior performance over GA and IMODE. Various mathematical tests and applications outside of cosmology support this \cite{Enriquez_Mendoza_Velasco_2022}.

The drop in performance of IMODE and GA can be attributed to their reliance on hyperparameters and operators that require fine-tuning for specific applications, whereas we used default hyperparameters based on their standard implementations. While such tuning could potentially improve performance, it often comes at the cost of significantly increased computational time and complexity.

To further investigate the performance of EAs, we analyzed their behavior for a single mock data realization of expansion rate, growth rate, and supernovae measurements, varying the hyperparameter tied to the maximum number of function evaluations (Max\text{-}FES). Max\text{-}FES, determined by the product of the population size and the number of generations, limits the total number of objective function evaluations an EA can perform (Section \ref{sec:evolutionary_algorithm}). While a standard Max\text{-}FES value of 500 sufficed for accurate and precise parameter estimation with $H(z)$+$f\sigma_8(z)$ data, this setting proved inadequate for IMODE and GA with default settings when SNe data were added. Table \ref{tab:computing_times} presents the precision obtained by each method and the corresponding computational time for varying Max\text{-}FES values.

\begin{table}[h!]
    \centering
    \caption{Computation times (in \texttt{MATLAB}) for one realization/mock data and the resulting precision of GA, IMODE, and PEOA for cosmological parameter estimation for different maximum number of function evaluations (Max\text{-}FES) when expansion, growth, and supernovae data are considered. Machine specs used: {\bf Processor} 13th Gen Intel(R) Core(TM) i7-13700, 2100 Mhz, 16 Core(s), 24 Logical Processor(s) {\bf RAM} 16 GB.}
    \resizebox{\columnwidth}{!}{%
    \begin{tabular}{|c|c|ccc|c|}
    \hline
    Method & Max\text{-}FES & PREC[$\Omega_{m0}$] & PREC [$H_0$] & PREC[$\sigma_8$] & Time (s) \\ \hline
    \multirow{4}{*}{GA} &
    500 & $53\%$ & $4\%$ & $14\%$ & $381$ \\ 
    & 1000 & $53\%$ & $3\%$ & $14\%$ & $883$ \\ 
    & 2000 & $52\%$ & $3\%$ & $14\%$ & $2144$ \\ 
    & 5000 & $49\%$ & $3\%$ & $13\%$ & $4889$ \\ \hline 
    \multirow{4}{*}{IMODE} &
    500 & $40\%$ & $2\%$ & $11\%$ & $342$ \\ 
    & 1000 & $21\%$ & $1\%$ & $6\%$ & $715$ \\ 
    & 2000 & $10\%$ & $0.5\%$ & $3\%$ & $1394$ \\ 
    & 5000 & $9\%$ & $0.5\%$ & $3\%$ & $3504$ \\ \hline
    \multirow{4}{*}{PEOA} &
    500 & $9\%$ & $0.5\%$ & $3\%$ & $401$ \\ 
    & 1000 & $9\%$ & $0.5\%$ & $3\%$ & $804$ \\ 
    & 2000 & $9\%$ & $0.5\%$ & $3\%$ & $1258$ \\ 
    & 5000 & $9\%$ & $0.5\%$ & $3\%$ & $2840$ \\ \hline
    \end{tabular}
    }
    \label{tab:computing_times}
\end{table}

\begin{figure}[h!]
    \centering
    \includegraphics[width=0.485\textwidth]{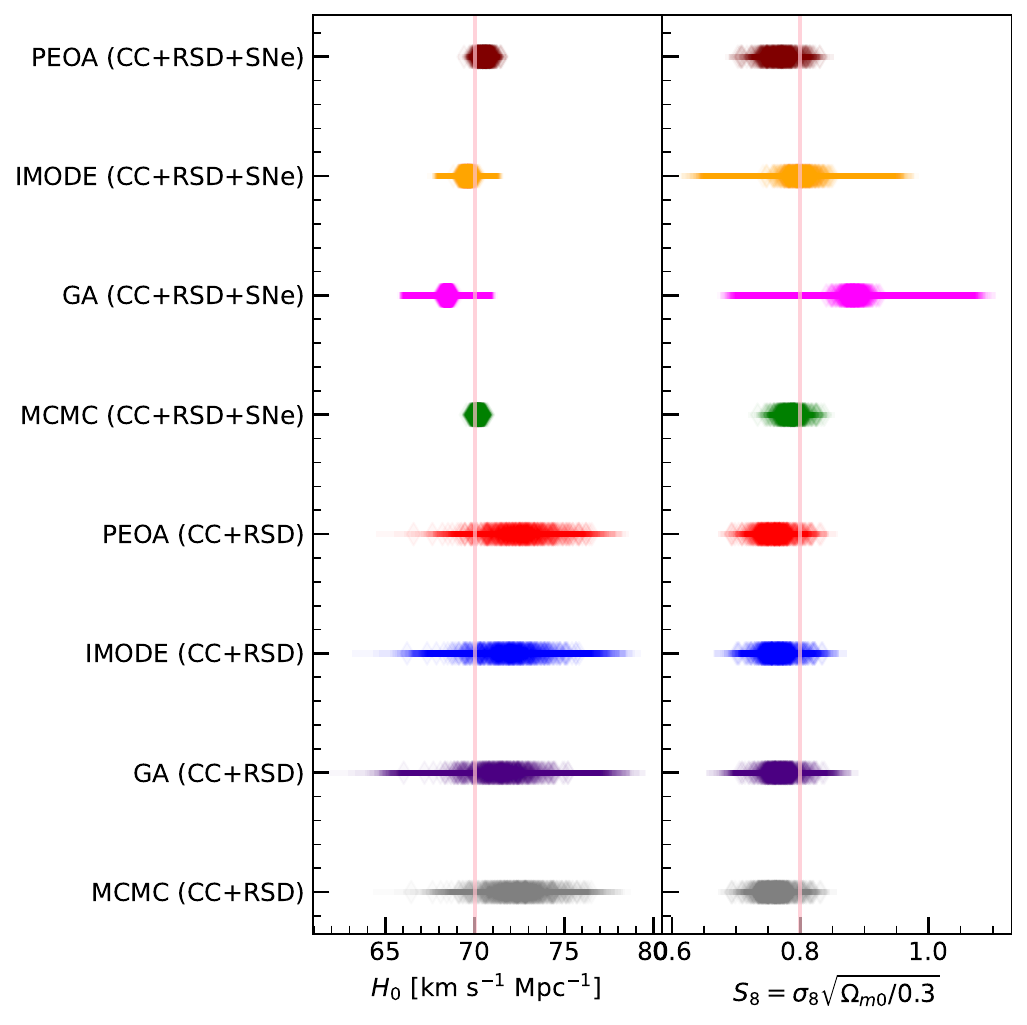}
    \caption{Whisker plot of $H_0$ and $S_8=\sigma_8 \sqrt{\Omega_{m0}/0.3}$ in each of 500 realizations of the mock data set considered. Pink vertical bands at $H_0=70$ km s Mpc$^{-1}$ and $S_8=0.8$ are the `true' cosmology.}
    \label{fig:H0S8_whisker}
\end{figure}

The results show that the performance drop observed in IMODE with the addition of supernovae data is primarily due to slow convergence. By increasing Max\text{-}FES from 500 to 2000, the previously wide parameter contours converged into visually acceptable distributions, relative to MCMC and PEOA, demonstrating the method’s potential with additional computational resources. However, this improvement comes with a cost: computational time increases roughly fourfold compared to the standard setup. GA, on the other hand, showed only marginal improvement in precision, even with Max\text{-}FES extended to 5000, which required nearly ten times the computational resources. This indicates that GA requires additional tuning of hyperparameters to improve its performance.

In contrast, PEOA displayed consistent and robust performance, even with default configuration. The algorithm's strong convergence properties and efficient exploration of the parameter space explain this. Appendix \ref{sec:convergence} presents an alternative way to view this discussion through the convergence toward a solution in each method.

Lastly, Figure \ref{fig:H0S8_whisker} shows the parameter estimates of $H_0$ and $S_8=\sigma_8 \sqrt{\Omega_{m0}/0.3}$ \cite{DiValentino:2020vvd, DiValentino:2025sru} obtained in each 500 realizations relative to the true cosmology in a whisker plot.

It is worth noting that the results of each method are consistent in all cases, though the constraints are not centered relative to the expected values\footnote{This can be addressed by including the covariance in the analysis, since our mock data sets are based on correlated observations that were treated independently.}. Regardless, the overall offset is consistent for all of the methods and for this work important observation is the performance of EA with respect to MCMC. The whisker plot supports the sentiments on accuracy and precision of EA that have been reiterated throughout the results and the discussion. {Note that when the full covariances of the data sets are considered, we can expect that generally the posteriors will be broader and in the case of very stringent data sets potential systematic biases on parameter estimation can be mitigated.}

\section{Conclusions}
\label{sec:conclusions}

In this work, we introduced two novel evolutionary algorithms (EAs) for cosmological parameter estimation: the Improved Multi-Operator Differential Evolution (IMODE) and the Philippine Eagle Optimization Algorithm (PEOA). Among these, PEOA demonstrated performance comparable to the standard Markov Chain Monte Carlo (MCMC) method while surpassing other EAs, such as IMODE and the genetic algorithm (GA). These results underscore the potential of PEOA in cosmological applications, although it is crucial to interpret such findings within the specific context of parameter estimation. In optimization problems, there is no universally superior method; thus, this study serves as an initial benchmark for exploring EAs in cosmology {with full systematics}.

We emphasize that EAs are fundamentally optimization algorithms, designed to find global optima in high dimensional, complex, and nonlinear spaces, but not reconstruct posterior probability distributions. Our results show that complemented by a general bootstrapping mechanism, EAs (beyond GA) are able to produce comparable output to MCMC by simply maximizing the likelihood, without relying on a full Bayesian framework. PEOA has turned out surprising in this light, giving competitive estimates of the cosmological parameters in this baseline study.

The motivation for exploring alternative statistical methods is not to replace existing techniques but to complement them. This is particularly relevant in cosmology, where diverse data sets---ranging from supernovae and the cosmic microwave background to baryon acoustic oscillations and gravitational waves---must be reconciled under a consistent physical framework. Similarly, the application of diverse statistical methods, including EAs, can provide robust cross-validation and novel perspectives.
{{Granted, the present work has no cosmological consequences and dealt with mock data. Nonetheless, it}} represents a step toward establishing a more diverse set of tools for confronting cosmological data.

Future research directions naturally emerge from this study. First, while GA has been applied to real cosmological data, the potential of IMODE and PEOA, among a full universe of nature-inspired EAs, remains to be tested on such data sets, which include full covariance structures. This will offer insights into their performance in more realistic scenarios. Additionally, EAs could play a significant role in cosmological model selection and in reconstructing cosmological functions such as spatial curvature and the dark energy density. Another promising avenue lies in leveraging the ability of EAs to explore high-dimensional and complex parameter spaces to investigate potential systematic effects in the data. {Through simplified mock data sets, this work has opened up a path to a field of biology-inspired optimization techniques (beside the few well knowns such as genetic algorithm and particle swarm optimization) for cosmology, but the simplicity has also concealed EAs full potential. Fully tapping into EAs mathematical capabilities such as exploring multi-modal functions or likelihoods remains a worthy consideration in a future work, in cosmological parameter estimation and in other astrophysical contexts.} On a practical side, we are preparing user-friendly \texttt{python} implementations of IMODE and PEOA to facilitate their adoption to cosmology and astrophysics.

\section*{Acknowledgements}
The authors are grateful to Isidro Gomez Vargas and Jackson Levi Said for their rigorous feedback on a preliminary version of the manuscript. This work was supported by the Office of the Vice Chancellor for Research and Development, University of the Philippines Diliman, Quezon City 1101, Philippines [Open Grant Project No. 242403 OG]. The authors recognize the networking support of the CosmoVerse COST Action (CA21136 - Addressing observational tensions in cosmology with systematics and fundamental physics). RCB is supported by an appointment to the JRG Program at the APCTP through the Science and Technology Promotion Fund and Lottery Fund of the Korean Government, and was also supported by the Korean Local Governments in Gyeongsangbuk-do Province and Pohang City. RCB thanks the Institute of Physics, Academia Sinica for the kind hospitality that enabled the completion of this work. RR was affiliated with the National Institute of Physics, University of the Philippines Diliman, during the conceptualization of this work until the completion of a preliminary version of the manuscript and is currently affiliated with the Philippine Space Agency and the Research Center for Theoretical Physics, Central Visayan Institute Foundation. RR wishes to thank all of these institutions for the academic freedom and invaluable support they provided that made the conduct and completion of this work possible.


\onecolumn
\appendix

\section{Pseudo-algorithms}
\label{sec:pseudo-algorithms}
In this section, we present the  pseudo-codes for the three considered algorithms: GA, IMODE, and PEOA. Firstly, we define the common parameters for the three algorithms.
\begin{multicols}{2}
\begin{itemize}
    \item $f$: objective function
    \item $[ X_{\textrm{min}}, X_{\textrm{max}}]$: search space
    \item $Max\text{-}FES$: maximum function evaluations
    \item $D$: dimension
    \item $x^*$: optimal solution
    \item $f^*$: function value of $x^*$ \,.
\end{itemize}
\end{multicols}

\subsection{Genetic Algorithm}
\label{subsec:ga_pseudocode}
The basic steps of GA are given in the pseudocode displayed in Algorithm \ref{algo:ga}. The parameters specific in GA are the following:
\begin{multicols}{2}
\begin{itemize}
    \item $PS$: population size
    \item $cr$: crossover rate
    \item $m$: size of elite pool
\end{itemize}
\end{multicols}

\begin{algorithm}[h!]
\caption{Genetic Algorithm}
\begin{algorithmic}[1]
\renewcommand{\algorithmicrequire}{\textbf{Input:}}
\renewcommand{\algorithmicensure}{\textbf{Output:}}
\REQUIRE $f$, $D$, $PS$, lower bound $\mathbf{X}_{\min}$, upper bound $\mathbf{X}_{\max}$, $cr$, $m$
\ENSURE $f(\mathbf{x}^*)$, $\mathbf{x}^*$
\STATE Generate initial random population of size $PS$.;
\WHILE{$FuncEvals < Max\text{-}FES$}
    \STATE Rank all the individuals in the population by their cost function value and choose the best $m$ individuals to form the elite pool.
    \STATE Apply a tournament selection with size $TC$ and fill the selection pool. 
	\STATE Generate a random number $\bar{u}$ in $[0,1]$. 
	\FOR{each two consecutive individuals in the selection pool}
	    \IF{one of the selected individual is the same to another}
	        \STATE Replace one by a randomly-selected individual in the selection pool.
	    \ENDIF
	    \IF{$\bar{u} < cr$}
	        \STATE Generate an offspring from the two parents.
		\ENDIF
	    \FOR{each offspring}
		    \STATE Mutate the offspring using an individual from the archive pool.
	    \ENDFOR
	\ENDFOR
    \STATE The new generation is the collection of the elites and the offspring while maintaining the population size $PS$.
\ENDWHILE
\RETURN fittest individual $\mathbf{x}^*$ and $f(\mathbf{x}^*)$
\end{algorithmic} \label{algo:ga}
\end{algorithm}

\clearpage

\subsection{Improved Multi-Operator Differential Evolution Algorithm}
\label{subsec:imode_pseudocode}

The basic steps of IMODE are given in the pseudocode displayed in Algorithm \ref{algo2}. We have the following parameters for IMODE:

\begin{multicols}{2}
\begin{itemize}
    \item $Prob_{ls}$: 
    \item $prob_1$: 
    \item $prob_2$: 
    \item $NP$: 
    \item $NP_{op}$: 
    \item $QR_{op}$: 
\end{itemize}
\end{multicols}

\begin{algorithm}[h!]
\caption{IMODE Algorithm}
\begin{algorithmic}[1]
\STATE Define $Prob_{ls} \leftarrow 0.1$, $Max\text{-}FES$, $prob_1 \leftarrow 1$, $prob_2 \leftarrow 1$, $NP$, $G \leftarrow 1$, and $FES \leftarrow 0$;
\STATE Generate an initial random population $(X)$ of size $N P$;
\STATE Evaluate $f(X)$, and update number of fitness evaluations    $F E S \leftarrow F E S+N P ;$
\STATE Each operator $o p$ is assigned the same number of solutions    $N P_{o p}$;
\WHILE{ $F E S \leq Max\text{-}FES$\ }
\STATE $G \leftarrow G+1 ;$
\STATE Generate new population using the assigned DE operators, i.e., each operator $o p$ evolves its assigned number of
    individuals $N P_{o p}$;
\STATE Calculate the diversity obtained from each operator    $o p$ $\left(D_{o p}\right)$ and the quality rate of solutions $Q R_{o p}$;
\STATE Update the number of solutions ($N P_{o p}$) each DE operator evolves;
\STATE Generate new population using the assigned DE operators;
\STATE Evaluate $f(X)$, and update number of fitness evaluations $F E S \leftarrow F E S+N P$;
\STATE Update $N P$ 2;
\IF{$F E S \geq 0.85 \times Max\text{-}FES$}
\STATE Apply local search;
\STATE Update FES
\ENDIF
\ENDWHILE
\end{algorithmic} \label{algo2}
\end{algorithm}


\clearpage

\subsection{Philippine Eagle Optimization Algorithm}
\label{subsec:peoa_pseudocode}

The basic steps of PEOA are summarized in the pseudocode shown in Algorithm \ref{algo1}, which uses the following variables:

\begin{multicols}{2}
\begin{itemize}
    \item $S_0$: initial eagle population size
    \item $S_{\textrm{loc}}$: local food size
    \item $K$: generation number
    \item $N$: function evaluations counter
    \item $P_i$: probability for operator $i$, $i=1,2,3$
    \item $P$: probability vector $P= [P_1,P_2,P_3]$
    \item $X$: initial population of eagles
    \item $X^*$: best eagle in $X$
    \item $Y^*$: best searched food of $X^*$
    \item $X_{\textrm{new}}$: new population of eagles
    \item $f_{\textrm{true}}$: true optimal value
\end{itemize}
\end{multicols}

\begin{algorithm}[H]
\caption{Philippine Eagle Optimization Algorithm}
\begin{algorithmic}[1]
\renewcommand{\algorithmicrequire}{\textbf{Input:}}
\renewcommand{\algorithmicensure}{\textbf{Output:}}
\REQUIRE $f$, $X_{\textrm{min}}$, $X_{\textrm{max}}$, $D$
\ENSURE $x^*$, $f^*$
\STATE Define $Max\text{-}FES$, $S_0$, and $S_{\textrm{loc}}$.
\STATE Set $K \leftarrow 0$, $N \leftarrow 0$, and for each $i = 1, 2, 3$, $P_i \leftarrow \tfrac{1}{3}$. \\
\COMMENT{\texttt{Initialization Phase}}
\STATE Generate initial population of eagles $X$ of size $S_0$.
\STATE Sort $X$ based on function value {to determine $X^\star$} and update $N$. \\
\COMMENT{\texttt{Local Phase}}
\STATE {Search $Y^\star$}
via interior point method with maximum evaluations $S_{\textrm{loc}}$ and update $N$.
\WHILE{ $|f(Y^\star) - f_{\textrm{true}}| \geq 10^{-8}$ \textbf{or} $N \leq Max\text{-}FES$\ }
\STATE Set $K = K + 1$.
\STATE Update $S$ via linear population size reduction.
\STATE Divide eagle population into subpopulations using $P$. \\
\COMMENT{\texttt{Global Phase}}
\STATE Generate new population of eagles $X_{\textrm{new}}$ via the corresponding operators assigned to the subpopulations.
\STATE Sort $X_{\textrm{new}}$ based on function value {to obtain the new $X^\star$} and update $N$. \\
\COMMENT{\texttt{Local Phase}}
\STATE {Repeat the Local Phase (Step 5) with the updated $X^\star$}.
\STATE Update $P$ based on the improvement rate of each operator.
\ENDWHILE
\RETURN $x^* = Y^\star$ and $f^* = f(Y^\star)$
\end{algorithmic} \label{algo1}
\end{algorithm}


\newpage
\twocolumn
\section{Convergence}
\label{sec:convergence}

\begin{figure}[h!]
    \centering
    \includegraphics[width=0.49\textwidth]{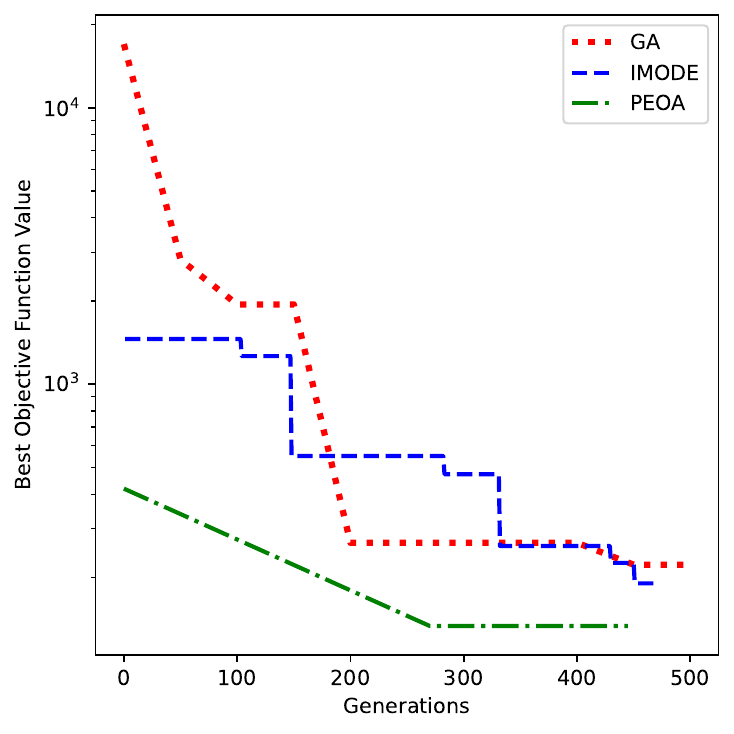}
    \caption{Convergence in a sample run of GA, IMODE and PEOA with standard configurations/hyperparameter setup. One realization of expansion rate, growth rate, and supernovae mock data is used.}
    \label{fig:convergence}
\end{figure}

Complementing the discussion on convergence toward a solution (Section \ref{sec:discussion}), presented through computation times, we show the evolution of the optimized objective function in each method for a single run in a single realization of mock data in Figure \ref{fig:convergence}.

The important observation is how fast a method approaches its final value. This shows clearly that while PEOA smoothly converges to a solution, both IMODE and GA take more generations to do so. Moreover, the fast convergence of PEOA is supported by a smaller objective function value, in contrast with the two other EAs.


\end{document}